\documentclass{emulateapj}

\newcommand{\iso}{{\em ISO}}

\newcommand{\mum}{\ifmmode{\rm \mu m}\else{$\mu$m}\fi}
\newcommand{\msol}{\mbox{M$_{\odot}$}}
\newcommand{\mdot}{$\dot{M}$}

\journalinfo{The Astrophysical Journal, scheduled for 718: in press, 
  2010 August 1}
\submitted{Received 2009 November 16; accepted 2010 June 18}

\begin{document}

\title{{\it Spitzer} spectroscopy of mass loss and dust
production by evolved stars in globular clusters}

\author{
G.~C.~Sloan\altaffilmark{1},
N.~Matsunaga\altaffilmark{2,3},
M.~Matsuura\altaffilmark{4,5},
A.~A.~Zijlstra\altaffilmark{6},
K.~E.~Kraemer\altaffilmark{7},
P.~R.~Wood\altaffilmark{8}
J.~Nieusma\altaffilmark{9,10,11},
J.~Bernard-Salas\altaffilmark{1},
D.~Devost\altaffilmark{12},
J.~R.~Houck\altaffilmark{1}
}
\altaffiltext{1}{Cornell University, Astronomy Department,
  Ithaca, NY 14853-6801, sloan@isc.astro.cornell.edu
  jbs@isc.astro.cornell.edu, jrh13@cornell.edu}
\altaffiltext{2}{Department of Astronomy, Kyoto University,
  Kitashirakawa-Oiwake-cho, Sakyo-ku, Kyoto 606-8502, Japan}
\altaffiltext{3}{Institute of Astronomy, University of Tokyo, 2-21-1
  Osawa, Mitaka, Tokyo 181-0015, Japan, matsunaga@ioa.s.u-tokyo.ac.jp}
\altaffiltext{4}{University College London -- Institute of Origins, 
  Department of Physics and Astronomy, Gower Street, London WC1E 6BT, 
  UK, mikako@star.ucl.ac.uk}
\altaffiltext{5}{University College London -- Institute of Origins, 
  Mullard Space Science Laboratory, Holmbury St.\ Mary, Dorking, 
  Surrey RH5 6NT, UK}
%\altaffiltext{4}{National Optical Astronomical Observatory of Japan,
%  Osawa 2-21-1, Mitaka, Tokyo 181-8588, Japan}
\altaffiltext{6}{University of Manchester, School of Physics \& Astronomy,
   P.~O.~Box 88, Manchester M60 1QD, UK, albert.zijlstra@manchester.ac.uk}
\altaffiltext{7}{Air Force Research Laboratory, Space Vehicles Directorate,
  Hanscom AFB, MA 01731}
\altaffiltext{8}{Research School of Astronomy and Astrophysics,
  Australian National University, Cotter Road, Weston Creek ACT 2611,
  Australia, wood@mso.anu.edu.au}
\altaffiltext{9}{Department of Physics, The College of New Jersey,
  2000 Pennington Rd., Ewing, NJ 08628}
\altaffiltext{10}{Department of Astronomy, University of Michigan,
  500 Church St., Ann Arbor, MI 48109, judaniel@umich.edu}
\altaffiltext{11}{NSF REU Research Assistant, Cornell University,
  Astronomy Department, Ithaca, NY 14853-6801}
\altaffiltext{12}{Canada France Hawaii Telescope, 65-1238 Mamalahoa 
  Hwy, Kamuela, HI, 96743, devost@cfht.hawaii.edu} 

\begin{abstract}

We have observed a sample of 35 long-period variables and 
four Cepheid variables in the vicinity of 23 Galactic 
globular clusters using the Infrared Spectrograph on the 
{\it Spitzer Space Telescope}.  The long-period variables
in the sample cover a range of metallicities from near solar
to about 1/40th solar.  The dust mass-loss rate from the 
stars increases with pulsation period and bolometric 
luminosity.  Higher mass-loss rates are associated with 
greater contributions from silicate grains.  The dust
mass-loss rate also depends on metallicity.  The dependence
is most clear when segregating the sample by dust 
composition, less clear when segregating by bolometric 
magnitude, and absent when segregating by period.  The 
spectra are rich in solid-state and molecular features.  
Emission from alumina dust is apparent across the 
range of metallicities.  Spectra with a 13-\mum\ dust emission 
feature, as well as an associated feature at 20~\mum, also
appear at most metallicities.  Molecular features in the
spectra include H$_2$O bands at 6.4--6.8~\mum, seen in both
emission and absorption, SO$_2$ absorption at 7.3--7.5~\mum, 
and narrow emission bands from CO$_2$ from 13.5 to 16.8~\mum.
The star Lyng{\aa}~7~V1 has an infrared spectrum revealing it 
to be a carbon star, adding to the small number of carbon 
stars associated with Galactic globular clusters.
\end{abstract}


\section{Introduction} % Sec. 1

Stars ascending the asymptotic giant branch (AGB) burn
hydrogen and helium in shells around an inert C-O core 
\citep[e.g.,][]{ir81}.  The fusion of helium to carbon
proceeds by the triple-$\alpha$ sequence \citep{sal52} in
thermal pulses which lead to the dredge-up of freshly
produced carbon to the surface of the star.  If the envelope
of the star is not too massive and the dredge-ups are
sufficiently strong, then enough carbon reaches the surface 
to drive the photospheric C/O ratio over unity.  Generally, 
the formation of CO molecules will exhaust whichever of the 
carbon or oxygen is less abundant, leading to a chemical 
dichotomy.  Carbon stars produce carbon-rich dust, and 
oxygen-rich stars produce oxygen-rich dust.  Stars in the 
range from $\sim$2--5~\msol\ become carbon stars, although 
both the lower and upper mass limits decrease in more
metal-poor environments \citep[e.g.][]{kl07}.  

Several surveys with the Infrared Spectrograph 
\citep[IRS;][]{hou04} on the {\it Spitzer Space Telescope} 
\citep{wer04} have probed how the production of dust by 
evolved stars depends on metallicity by observing AGB stars 
and supergiants in the Magallanic Clouds and other nearby 
Local Group galaxies.  Carbon stars dominate these samples.  
The rate at which they produce dust does not vary 
significantly with metallicity \citep[][and references 
therein]{gro07,slo08}.  It would appear that they produce 
and dredge up all of the carbon they need to form dust, 
regardless of their initial abundances \citep{slo09}.  

The published Local Group samples contain fewer oxygen-rich 
evolved stars, making any conclusions about this population 
more tentative.  \cite{slo08} compared oxygen-rich sources in 
the Galaxy, Large Magellanic Cloud (LMC), and Small 
Magellanic Cloud (SMC), and they found that as the 
metallicity of the sample decreased, the fraction of stars 
with a dust excess decreased.  Infrared photometric surveys
of several dwarf irregular galaxies in the Local Group also
show a trend of less dust at lower metallicities 
\citep{boy09a}.

The spectroscopic evidence for a dependency of dust 
production on metallicity is not strong.  Stars in the 
Magellanic Clouds can pulsate with periods of $\sim$700 days 
or longer, and these longer-period variables are usually 
embedded in significant amounts of dust, with no apparent 
dependence on metallicity.  Additional oxygen-rich sources 
from the SMC further blur the trends noticed before 
\citep{slo10}.  One major problem with these samples is that 
oxygen-rich evolved stars can come from three distinct 
populations.   The red supergiants and intermediate-mass AGB 
stars (or super-AGB stars) are too massive to become carbon 
stars, while the low-mass AGB stars have too little mass.  
This mixing of populations confuses the observed samples of 
oxygen-rich evolved stars in the Galaxy, the Magellanic 
Clouds, and more distant irregulars in the Local Group.

% \citet{Dupree09} argued that mass-loss rate from red giants do 
% not depend on the metallicities.  This suggests that the 
% mass-loss process is not driven by radiation pressure on dust 
% in red giants in metal poor field stars.  In contrast, amongst 
% stars with dust driven wind, mass-loss rate appears to reduces 
% with lower metallicity \citep{Marshall04}.  This has been 
% observed using OH masers amongst OH/IR stars in the LMC.  This 
% shows metallicity dependence of mass-loss rate remain largely 
% uncertain, so as the mass loss driven mechanisms.

Globular clusters provide another means of investigating
dependencies of dust production on metallicity.  Like the
Magellanic Clouds, most globular clusters have known
metallicities, and they are at known distances, allowing us 
to directly determine their luminosities.  Unlike the 
Magellanic Clouds, globular clusters have old populations of 
stars, with few significantly younger than 10 billion years.  
This limits the sample to masses of $\sim$1\msol\ or less, 
below the lower mass limit for carbon stars.

\cite{leb06} identified and observed 11 long-period variables 
(LPVs) in the globular cluster 47 Tuc with the IRS on {\it 
Spitzer}.  Their data were consistent with a shift from 
relatively dust-free AGB stars to more deeply embedded 
sources with higher mass-loss rates at a luminosity of 
$\sim$2000~L$_{\sun}$.  But with a sample of only one 
globular cluster, they were unable to address the important 
question of how the dust properties depend on metallicity.

% MORE OUTLINE:  Could discuss
% Woitke, H\"{o}fner, Mattson, Lagadec and Zijlstra

We have used the IRS on {\it Spitzer} to observe a sample of 
39 variable stars in 23 globular clusters spanning a range of 
metallicity from nearly solar to only a few percent of solar.
This paper presents an overview of the program.
\S~\ref{SecObs} describes the sample of clusters and 
individual stars and the observations, including 
near-infrared photometric monitoring prior to the {\it 
Spitzer} observations, the IRS spectroscopy, and nearly
simultaneous near-infrared photometry.  \S~\ref{SecMem} 
assesses the membership of our targets within the clusters,
while \S~\ref{SecAnal} describes the analysis and the 
results.  \S~\ref{SecObject} treats some unusual objects 
individually, and \S~\ref{SecEvo} examines the larger picture 
of evolution on the AGB.

\section{Observations} % Sec. 2
\label{SecObs} 

\begin{deluxetable*}{llrlllll} % Table 1
%\begin{deluxetable}{llrlllll} 
%\rotate
\tablecolumns{8}
\tablewidth{0pt}
\tablecaption{Globular clusters in the study\label{TblClust}}
\tablehead{
  \colhead{Globular} & \colhead{Alternative} & \colhead{} & \colhead{} & 
  \colhead{} & \colhead{} & \colhead{Distance} & \colhead{} \\
  \colhead{Cluster} & \colhead{Name} & \colhead{[Fe/H]} & 
  \colhead{Ref.\ \tablenotemark{a}} & \colhead{E(B$-$V)} & \colhead{Ref.\ \tablenotemark{a}} & \colhead{Modulus} & 
  \colhead{Ref.\ \tablenotemark{a}} \\
}
\startdata

NGC 362     &              & $-$1.20 & 1,2,3,4,5,6  & 0.04 $\pm$ 0.02                  & 1,3,4,7    & 14.83 $\pm$ 0.15 & 1,3,4,6,7,8 \\
NGC 5139    & $\omega$ Cen & $-$1.63 & 1,5          & 0.12 $\pm$ 0.01\tablenotemark{b} & 1          & 13.70 $\pm$ 0.17 & 1,9,10,11,12,13,14,15,16 \\
NGC 5904    & M 5          & $-$1.20 & 2,3,4,5,6    & 0.03 $\pm$ 0.01                  & 1,3,4,7,17 & 14.46 $\pm$ 0.10 & 1,3,4,6,7,8,17 \\
NGC 5927    &              & $-$0.35 & 1,3,4,5,6    & 0.46 $\pm$ 0.05\tablenotemark{b} & 1,3        & 14.45 $\pm$ 0.07 & 1,3,4,6 \\
Lyng{\aa} 7 &              & $-$0.66 & 1,6,18       & 0.73 $\pm$ 0.12                  & 1,18       & 14.31 $\pm$ 0.10 & 1,18 \\
NGC 6171    & M 107        & $-$0.98 & 1,3,4,5,6    & 0.38 $\pm$ 0.08                  & 1,3,4,17   & 13.89 $\pm$ 0.18 & 1,3,4,17 \\
NGC 6254    & M 10         & $-$1.51 & 1,3,5,6      & 0.28 $\pm$ 0.03\tablenotemark{b} & 1,3        & 13.30 $\pm$ 0.12 & 1,3 \\
NGC 6352    &              & $-$0.69 & 1,3,5,6      & 0.21 $\pm$ 0.02\tablenotemark{b} & 1,3        & 13.85 $\pm$ 0.06 & 1,3,17 \\
NGC 6356    &              & $-$0.50 & 1,4,5        & 0.28 $\pm$ 0.03\tablenotemark{b} & 1          & 15.93 $\pm$ 0.09 & 1,4 \\
NGC 6388    &              & $-$0.57 & 1,4,5,19     & 0.35 $\pm$ 0.04                  & 1,20       & 15.47 $\pm$ 0.12 & 4,20,21 \\
Palomar 6   &              & \nodata & 22           & 1.36 $\pm$ 0.14\tablenotemark{b} & 1,23,24,25 & 14.23 $\pm$ 0.40 & 1,23,24,25 \\
Terzan 5    &              & $-$0.08 & 1,5,23,26,27 & 2.31 $\pm$ 0.23\tablenotemark{b} & 1,23,27    & 14.11 $\pm$ 0.13 & 28 \\
NGC 6441    &              & $-$0.56 & 1,4,5,29     & 0.50 $\pm$ 0.05\tablenotemark{b} & 1,29       & 15.57 $\pm$ 0.16 & 1,4,29,30,31 \\
NGC 6553    &              & $-$0.28 & 1,3,5        & 0.72 $\pm$ 0.11                  & 1,3,23     & 13.63 $\pm$ 0.22 & 1,3,23,30 \\
IC 1276     & Palomar 7    & $-$0.69 & 1,5          & 1.12 $\pm$ 0.11\tablenotemark{b} & 1,32       & 13.66 $\pm$ 0.15\tablenotemark{b} & 1 \\
Terzan 12   &              & $-$0.50 & 23           & 2.06 $\pm$ 0.21\tablenotemark{b} & 23         & 13.38 $\pm$ 0.15\tablenotemark{b} & 1 \\
NGC 6626    & M 28         & $-$1.46 & 1,5          & 0.40 $\pm$ 0.04\tablenotemark{b} & 1          & 13.73 $\pm$ 0.15\tablenotemark{b} & 1 \\
NGC 6637    & M 69         & $-$0.66 & 1,3,4,5,6,33 & 0.16 $\pm$ 0.02                  & 1,3,33     & 14.76 $\pm$ 0.09 & 1,3,4,21,33 \\
NGC 6712    &              & $-$0.97 & 1,3,5        & 0.43 $\pm$ 0.04                  & 1,3,4      & 14.24 $\pm$ 0.09 & 1,3,4 \\
NGC 6760    &              & $-$0.48 & 1,4,5        & 0.77 $\pm$ 0.08\tablenotemark{b} & 1          & 14.59 $\pm$ 0.29 & 1,4 \\
NGC 6779    & M 56         & $-$2.05 & 1,5,6        & 0.19 $\pm$ 0.08                  & 1,34       & 15.23 $\pm$ 0.30 & 1,34 \\
Palomar 10  &              & $-$0.10 & 1            & 1.66 $\pm$ 0.17\tablenotemark{b} & 1,31       & 13.86 $\pm$ 0.37 & 1,32 \\
NGC 6838    & M 71         & $-$0.73 & 1,2,3,4,5,6  & 0.26 $\pm$ 0.03\tablenotemark{b} & 1,2,3      & 12.99 $\pm$ 0.10 & 1,3,4,7 \\
\enddata
\tablenotetext{a}{References:
(1) \citet{har03}; (2) \citet{cg97}; (3) \citet{fer99}; (4) \citet{rb05};
(5) \citet{car09}; (6) \citet{dot10}; (7) \citet{car00}; (8) \citet{gra97}; 
(9) \citet{md00} ; (10) \citet{mcn00}; (11) \citet{tho01}; (12) \citet{kal02};
(13) \citet{cap02}; (14) \citet{dp06}; (15) \citet{cat06}; (16) \citet{vdv06}; 
(17) \citet{sw98}; (18) \citet{sar04}; (19) \citet{car07}; (20) \citet{dal08};
(21) \cite{mat07b}; (22) See \S~\ref{SecClust}; (23) \citet{bar98a}; 
(24) \citet{ort98}; (25) \citet{lc02}; (26) \citet{or04}; (27) \citet{val07}; 
(28) This work (\S~\ref{SecPK}); (29) \citet{val04}; (30) \citet{hr99}; 
(31) \cite{mat09}; (32) \cite{bar98b}; (33) \citet{val05}; (34) \citet{iva00}; 
(35) \citet{kai97}.}
\tablenotetext{b}{Uncertainty assumed.}
\end{deluxetable*}
%\end{deluxetable}

\begin{deluxetable*}{llllllrrrrrrrl} % Table 2
%\begin{deluxetable}{llllllrlrrrlrl}
%\rotate
\tabletypesize{\small}
\tablecolumns{14}
\tablewidth{0pt}
\tablecaption{Observed variables in globular clusters\label{TblVar}}
\tablehead{
  \colhead{} & \colhead{RA} & \colhead{Dec.} & 
  \multicolumn{2}{c}{IRS observations} & \colhead{Var.} & 
  \colhead{Period} & \multicolumn{3}{c}{Mean magnitudes} & \colhead{} & 
  \multicolumn{2}{c}{Phase} & \colhead{} \\
  \colhead{Target} & \multicolumn{2}{c}{J2000} & \colhead{AOR key} & 
  \colhead{Date (JD)} & \colhead{Class\tablenotemark{a}} & \colhead{(days)} & 
  \colhead{$<$J$>$} & \colhead{$<$H$>$} & \colhead{$<$K$_s>$} & 
  \colhead{$\Delta$K$_s$} & \colhead{Zero (JD)} & 
  \colhead{Obs.\tablenotemark{b}} & \colhead{Notes} \\
}
\startdata
NGC 362 V2     & 01 03 21.85 & $-$70 54 20.1 & 21740800 & 2454268 & SR      &  89.0 &  9.74   &  9.11   &  8.93 & 0.18    & 2452445.7 & 0.48 \\
NGC 362 V16    & 01 03 15.10 & $-$70 50 32.3 & 21740800 & 2454268 & Mira/SR & 138   &  9.09   &  8.41   &  8.18 & 0.52    & 2452495.5 & 0.81 \\
NGC 5139 V42   & 13 26 46.36 & $-$47 29 30.4 & 21741056 & 2454310 & Mira    & 149   &  8.25   &  7.52   &  7.14 & 0.52    & 2452525.7 & 0.97 \\
NGC 5904 V84   & 15 18 36.15 & $+$02 04 16.3 & 21741312 & 2454316 & CW      &  25.8 & 10.05   &  9.67   &  9.60 & 0.69    & 2452400.5 & \nodata \\
NGC 5927 V1    & 15 28 15.17 & $-$50 38 09.3 & 21741568 & 2454348 & Mira/SR & 202   &  9.17   &  8.27   &  7.87 & 0.38    & 2452475.6 & 0.29 \\
NGC 5927 V3    & 15 28 00.13 & $-$50 40 24.6 & 21741824 & 2454348 & Mira    & 297   &  8.72   &  7.84   &  7.33 & 0.89    & 2452582.5 & 0.94 \\
Lyng{\aa} 7 V1 & 16 11 02.05 & $-$55 19 13.5 & 21742080 & 2454350 & Mira    & 551   & 11.35   &  9.08   &  7.25 & 1.22    & 2452856.2 & 0.71 \\
NGC 6171 V1    & 16 32 24.61 & $-$13 12 01.3 & 21742336 & 2454345 & Mira    & 332   &  6.02   &  5.16   &  4.54 & \nodata & \nodata   & \nodata & c \\
NGC 6254 V2    & 16 57 11.74 & $-$04 03 59.7 & 21742592 & 2454347 & CW      &  19.7 & 10.12   &  9.59   &  9.44 & 0.71    & 2452394.4 & \nodata \\
NGC 6352 V5    & 17 25 37.52 & $-$48 22 10.0 & 21742848 & 2454384 & Mira    & 177   &  8.34   &  7.43   &  7.07 & 0.32    & 2452492.7 & 0.66 \\
NGC 6356 V1    & 17 23 33.72 & $-$17 49 14.8 & 21743104 & 2454375 & Mira    & 227   & 10.18   &  9.29   &  8.83 & 0.61    & 2452525.5 & 0.15 \\
NGC 6356 V3    & 17 23 33.30 & $-$17 48 07.4 & 21743104 & 2454375 & Mira    & 223   & 10.18   &  9.37   &  8.93 & 0.80    & 2452548.8 & 0.21 \\
NGC 6356 V4    & 17 23 48.00 & $-$17 48 04.5 & 21743104 & 2454375 & Mira    & 211   & 10.33   &  9.47   &  9.05 & 0.66    & 2452546.6 & 0.68 \\
NGC 6356 V5    & 17 23 17.06 & $-$17 46 24.5 & 21743104 & 2454375 & Mira    & 220   &  9.33   &  9.06   &  8.62 & \nodata & \nodata   & \nodata & c \\
NGC 6388 V3    & 17 36 15.04 & $-$44 43 32.5 & 21743360 & 2454377 & Mira    & 156   & 10.29   &  9.25   &  8.97 & 0.28    & 2452345.7 & 0.99 \\
NGC 6388 V4    & 17 35 58.94 & $-$44 43 39.8 & 21743360 & 2454377 & Mira    & 253   &  9.64   &  8.89   &  8.47 & 0.80    & 2452437.7 & 0.66 \\
Palomar 6 V1   & 17 43 49.48 & $-$26 15 27.9 & 21743616 & 2454374 & Mira    & 566   & \nodata & 11.65   &  8.82 & 1.53    & 2452653.9 & 0.04 \\
Terzan 5 V2    & 17 47 59.46 & $-$24 47 17.6 & 21743872 & 2454374 & Mira    & 217   &  9.78   &  8.38   &  7.65 & 0.67    & 2452464.4 & 0.80 \\
Terzan 5 V5    & 17 48 03.40 & $-$24 46 42.0 & 21744128 & 2454374 & Mira    & 464   & 10.03   &  8.04   &  6.83 & 0.86    & 2452557.9 & 0.91 \\
Terzan 5 V6    & 17 48 09.25 & $-$24 47 06.3 & 21744128 & 2454374 & Mira    & 269   & 10.01   &  8.41   &  7.50 & 0.61    & 2452366.0 & 0.47 \\
Terzan 5 V7    & 17 47 54.33 & $-$24 49 54.6 & 21744128 & 2454374 & Mira    & 377   &  9.74   &  8.02   &  7.03 & 0.76    & 2452511.8 & 0.95 \\
Terzan 5 V8    & 17 48 07.18 & $-$24 46 26.6 & 21744128 & 2454374 & Mira    & 261   &  9.79   &  8.28   &  7.44 & 0.80    & 2452459.2 & 0.35 \\
Terzan 5 V9    & 17 48 11.86 & $-$24 50 17.1 & 21743872 & 2454374 & Mira    & 464   & 11.43   &  9.55   &  8.41 & 0.83    & 2452401.0 & 0.25 \\
NGC 6441 V1    & 17 50 17.09 & $-$37 03 49.7 & 21744384 & 2454577 & Mira    & 200   & 10.28   &  9.26   &  8.90 & 0.40    & 2452495.1 & 0.40 \\   
NGC 6441 V2    & 17 50 16.16 & $-$37 02 40.5 & 21744384 & 2454577 & Mira    & 145   & 10.50   &  9.57   &  9.23 & 0.52    & 2452372.0 & 0.18 \\   
NGC 6553 V4    & 18 09 18.84 & $-$25 54 35.8 & 21744640 & 2454373 & Mira    & 267   &  8.17   &  7.21   &  6.66 & 1.06    & 2452727.2 & 0.16 \\   
IC 1276 V1     & 18 10 51.55 & $-$07 10 54.5 & 21742080 & 2454376 & Mira    & 222   &  8.43   &  7.31   &  6.77 & 0.29    & 2452511.4 & 0.41 \\   
IC 1276 V3     & 18 10 50.79 & $-$07 13 49.1 & 21742080 & 2454376 & Mira    & 300   &  8.27   &  7.25   &  6.60 & 0.89    & 2452422.2 & 0.51 \\   
Terzan 12 V1   & 18 12 14.18 & $-$22 43 58.9 & 21745152 & 2454373 & Mira    & 458   &  8.87   &  7.21   &  6.23 & 1.02    & 2452586.6 & 0.90 \\   
NGC 6626 V17   & 18 24 35.84 & $-$24 53 15.8 & 21745408 & 2454375 & CW      &  48.6 &  9.38   &  8.76   &  8.61 & 0.51    & 2452405.2 & \nodata \\   
NGC 6637 V4    & 18 31 21.88 & $-$32 22 27.7 & 21745664 & 2454582 & Mira    & 200   &  8.97   &  8.26   &  7.86 & 0.58    & 2452378.1 & 0.00 \\   
NGC 6637 V5    & 18 31 23.44 & $-$32 20 49.5 & 21745664 & 2454582 & Mira    & 198   &  8.94   &  8.21   &  7.82 & 0.73    & 2452376.1 & 0.16 \\   
NGC 6712 V2    & 18 53 08.78 & $-$08 41 56.6 & 21745920 & 2454385 & SR      & 109   &  9.11   &  8.45   &  8.16 & 0.54    & 2452446.2 & 0.86 \\   
NGC 6712 V7    & 18 52 55.38 & $-$08 42 32.5 & 21745920 & 2454385 & Mira    & 193   &  8.69   &  7.95   &  7.55 & 0.55    & 2452525.4 & 0.64 \\   
NGC 6760 V3    & 19 11 14.31 & $+$01 01 46.6 & 21746176 & 2454381 & Mira    & 251   &  9.12   &  8.24   &  7.69 & 0.72    & 2452417.7 & 0.82 \\   
NGC 6760 V4    & 19 11 15.03 & $+$01 02 36.8 & 21746432 & 2454381 & Mira    & 226   &  9.39   &  8.55   &  7.96 & 0.94    & 2452486.6 & 0.40 \\   
NGC 6779 V6    & 19 16 35.78 & $+$30 11 38.8 & 21746688 & 2454269 & CW      &  44.9 & 10.68   & 10.23   & 10.09 & 0.66    & 2453214.4 & \nodata \\   
Palomar 10 V2  & 19 17 51.48 & $+$18 34 12.7 & 21746944 & 2454269 & Mira    & 393   &  6.79   &  5.58   &  4.87 & \nodata & \nodata   & \nodata & c \\ 
NGC 6838 V1    & 19 53 56.10 & $+$18 47 16.8 & 21746944 & 2454269 & Mira/SR & 179   &  7.59   &  6.85   &  6.57 & 0.32    & 2452421.6 & 0.34 \\
\enddata
\tablenotetext{a}{SR = semi-regular; CW = Cepheid of the W Vir type 
    (i.e.\ Pop. II).}
\tablenotetext{b}{Phase during the IRS observation for P $>$ 50 days.}
\tablenotetext{c}{Photometry from 2MASS, variability class and period from 
    \cite{cle01}.}
\end{deluxetable*}
%\end{deluxetable}

\subsection{The Clusters} % Sec. 2.1
\label{SecClust}

Table~\ref{TblClust} presents the metallicity, reddening, and 
distance modulus to the clusters in our sample.  The values 
for each are based on a review of the literature, starting 
with the catalog of \cite{har03}, which was published in 1996 
and updated on the web in 2003.  Most of the estimates for 
metallicity and reddening are unchanged between the two
editions for our target clusters, but most of the distances
have changed, usually by small amounts.  We have also 
included measurements of the metallicity and distance modulus 
published since 1996, taking care not to overweight references
included in the 2003 updates to the bibliography of the 
catalog.  We have averaged the measurements for metallicity, 
reddening, and distance modulus.  For the latter two, we also 
quote the standard deviation as the uncertainty, which 
propagates through the quantities we derive.  We take the 
minimum uncertainty for $E(B-V)$ to be 10\% of the mean.  In 
some cases, we have quoted the uncertainties given by the 
referenced papers.  Some data which vary substantially from 
the others have been excluded (and are not included in the 
references in Table~\ref{TblClust}).  

For IC~1276 and Terzan~12, we excluded distance modulus 
measurements inconsistent with our estimates based on the 
period-K relation (\S~\ref{SecPK}).  For Terzan~5, we report 
the distance modulus based on the analysis of the four 
relatively unreddened variables in our sample 
(\S~\ref{SecPK}).

The metallicity of Palomar~6 is problematic, with 
measurements in the refereed literature ranging from
$-$1.08 \citep{lc02} to +0.2 \citep{min95}.  Intermediate
measurements of $-$0.74 \citep{zin85} and $-$0.52 
\citep{sf04} have also been reported.  Averaging these
metallicities gives $<$[Fe/H]$>$ = $-$0.54$\pm$0.54.  With
a standard deviation as large as the mean, we are unable to
determine a metallicity for this cluster with any confidence.

The metallicity of the cluster is important, because we 
assign individual stars to bins based on their metallicity.
Each bin will contain between five and eight targets, all
with a similar metallicity.  By comparing how the properties
of the stars vary from one bin to the next, we can assess how 
the mass-loss and dust production depend on metallicity.

\subsection{Photometric Monitoring and Sample Selection} % Sec. 2.2
\label{SecSample}

\begin{figure*} % Fig. 1
%\includegraphics[width=6.5in]{figures/fphot.eps}
%\begin{figure}
\includegraphics[width=6.5in]{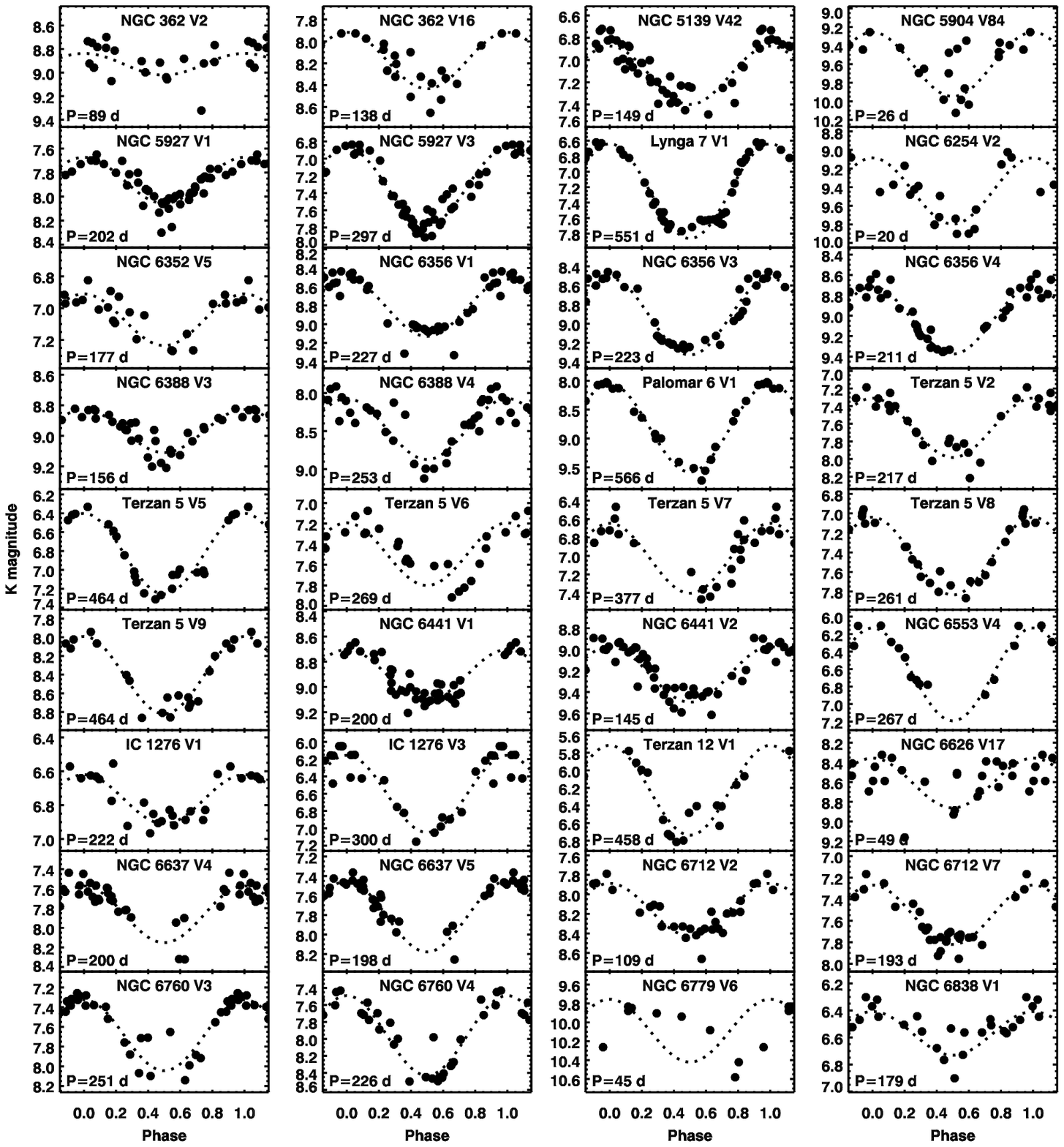}
\caption{Light curves in the K band for the sources in our 
sample observed with SIRIUS at the IRSF 1.4-m reflector.  
Table~\ref{TblVar} presents the fitted periods, mean 
magnitudes, K-band amplitude, and zero-phase dates.} 
\label{FigLC}
\end{figure*}
%\end{figure}

We selected the variable stars in the sample based on JHK 
photometry from the South African Astronomical Observatory 
(SAAO).  Observations were made using the SIRIUS 
near-infrared camera \citep[Simultaneous-Color Infrared 
Imager for Unbiased Surveys;][]{nag03} at the 1.4-m Infrared 
Survey Facility (IRSF) telescope over the period from 2002 
to 2005.  The IRSF/SIRIUS survey imaged fields in the 
vicinity of globular clusters at J, H, and K$_s$ (central 
wavelengths of 1.25, 1.63, and 2.14~\mum) with a typical 
sampling interval of 40--60 days \citep{mat07a}.  

The field of view of the survey was 7.7 arcmin$^2$, and most
clusters were imaged out to the half-mass radius.  Many new
variables were discovered, including eight new long-period
variables:  Lyng{\aa}~7~V1, Palomar~6~V1, Terzan~5~V5--V9, 
and Terzan~12~V1.  All eight of these stars are Miras, and
their periods range from 260 to 570 days.  The remaining
31 targets in our sample were previously known variables 
\citep[][and references therein]{cle01}. 

% \cite{mat05} reported
% masers in three of these new variables (Palomar~6~V1, 
% Terzan~5~V5, and Terzan~12~V1), as well as two previously
% known masers  (NGC~6171~V1 and Palomar~10~V2).

We determined periods, mean magnitudes, amplitudes and phases
for each variable by performing Fourier analysis and 
minimum-$\chi^2$ fits to the individual photometric 
observations in each SIRIUS filter for each target.  The 
results appear in Table~\ref{TblVar} and Figure~\ref{FigLC}.  
The quoted period and phase are the average from the three 
filters, except in cases where one filter disagreed 
substantially from the other two, or for Palomar~6~V1, which 
was too reddened for measurements at J.  Generally, the
uncertainty in the period was less than one day.\footnote{In
three cases, Lyng{\aa}~7~V1, Terzan~5~V9, and NGC~6838~V1, it 
was larger, 5--6 days.}  We lack monitoring observations of 
NGC~6171~V1, NGC~6356~V5, and Palomar~10~V2.  For these 
sources, Table~\ref{TblVar} contains variability information 
from \cite{cle01} and photometry from 2MASS 
\citep{skr06}.

The four ``CW'' variables are Pop.\ II Cepheids (of the W Vir
type).  They all have periods which are very short compared 
to the monitoring period, so any irregularities in the 
pulsations can accumulate to smear out the apparent 
periodicity in Figure~\ref{FigLC}.  Furthermore, we are 
unable to reproduce the published periods for NGC~6626~V17 
and NGC~6779~V6, which are 92.1 and 90 days, respectively 
\citep{cle01}.  Instead, we find a period of 48.6 days for 
NGC~6626~V17 and 44.9 days for NGC~6779~V6.  These appear to 
be overtones of the previously published periods, but we 
cannot say whether that is the result of an actual mode-shift 
in pulsation or sampling resolution of the older data.

\subsection{IRS Spectroscopy} % Sec. 2.3
\label{SecSpec}

\begin{figure} % Fig. 2
\includegraphics[width=3.5in]{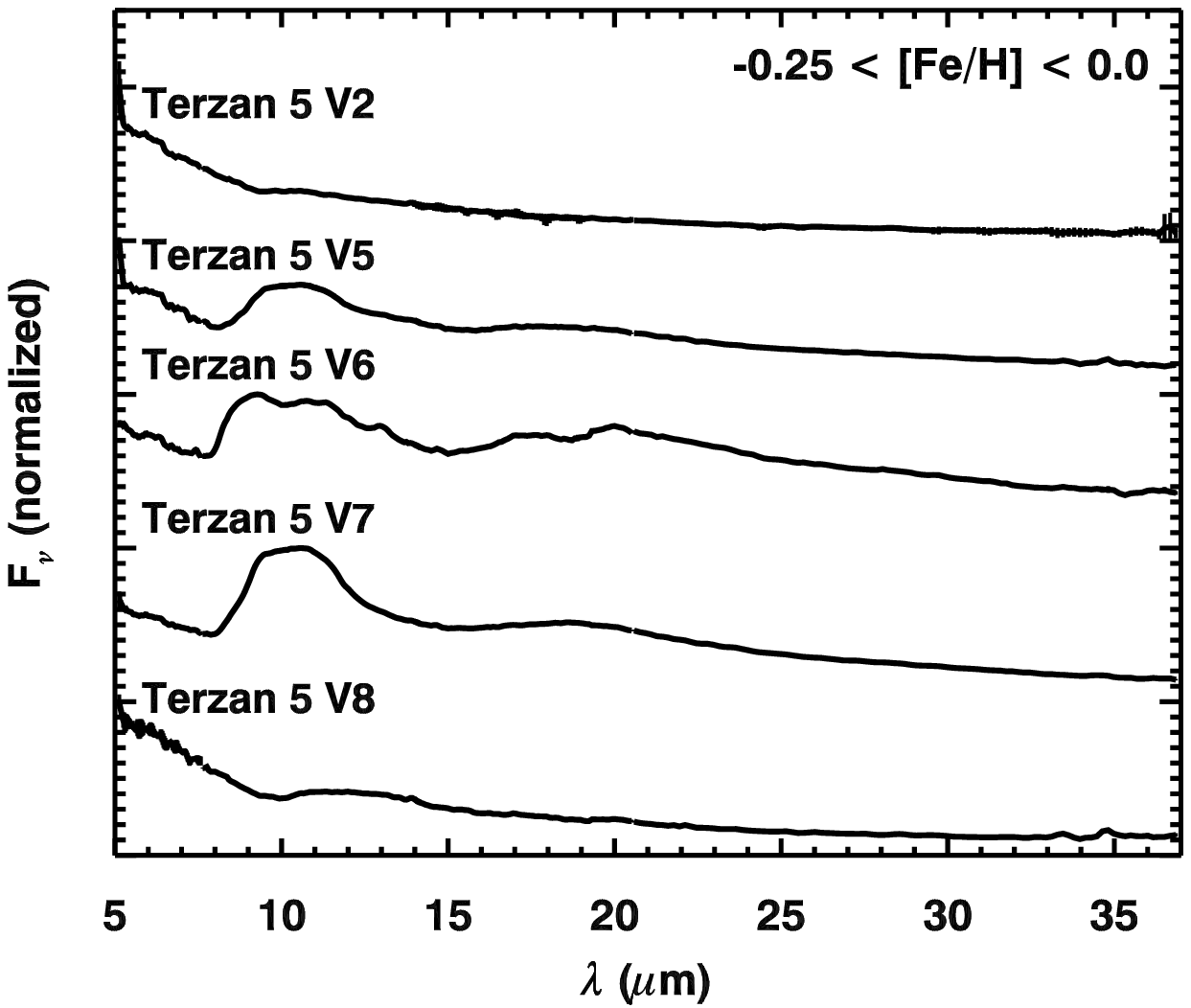}
\caption{IRS spectra of the five sources in our sample with
metallicities ([Fe/H]) between 0.0 and $-$0.25.  The error 
bars are generally smaller than the width of the plotted 
spectra.} 
\label{FigSp1}
\end{figure}

\begin{figure} % Fig. 3
\includegraphics[width=3.5in]{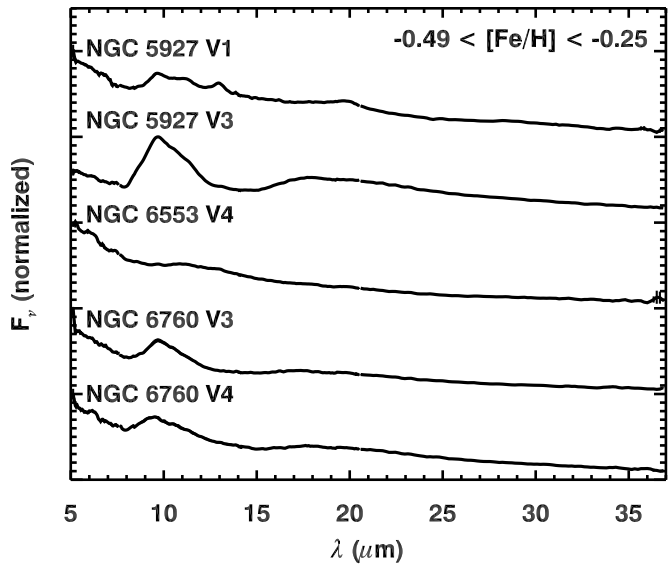}
\caption{IRS spectra of the five sources in our sample with
metallicities ([Fe/H]) between $-$0.25 and $-$0.49.} 
\label{FigSp2}
\end{figure}

\begin{figure} % Fig. 4
\includegraphics[width=3.5in]{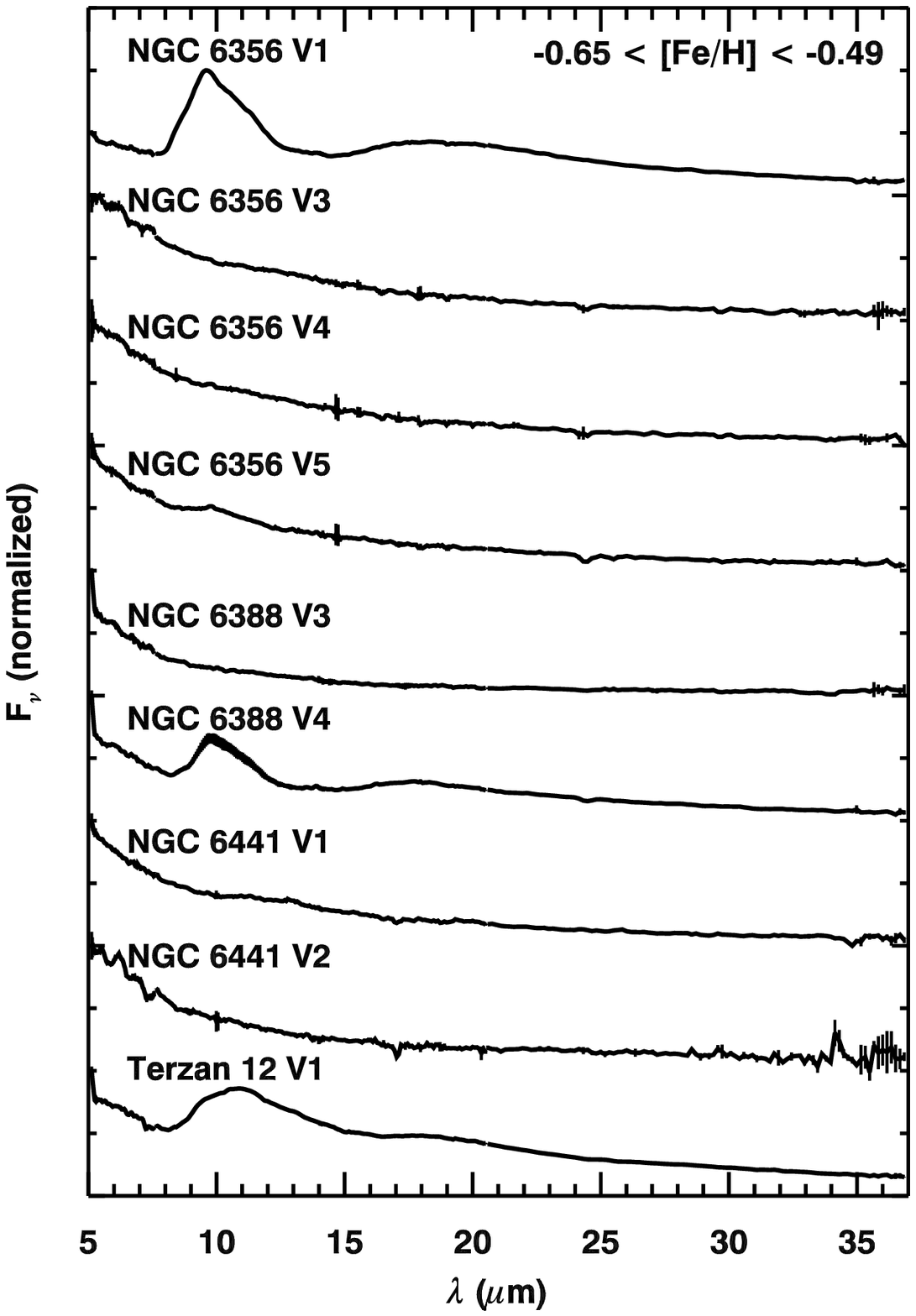}
\caption{IRS spectra of the nine sources in our sample with
metallicities ([Fe/H]) between $-$0.49 and $-$0.65.}
\label{FigSp3}
\end{figure} 

\begin{figure} % Fig. 5
\includegraphics[width=3.5in]{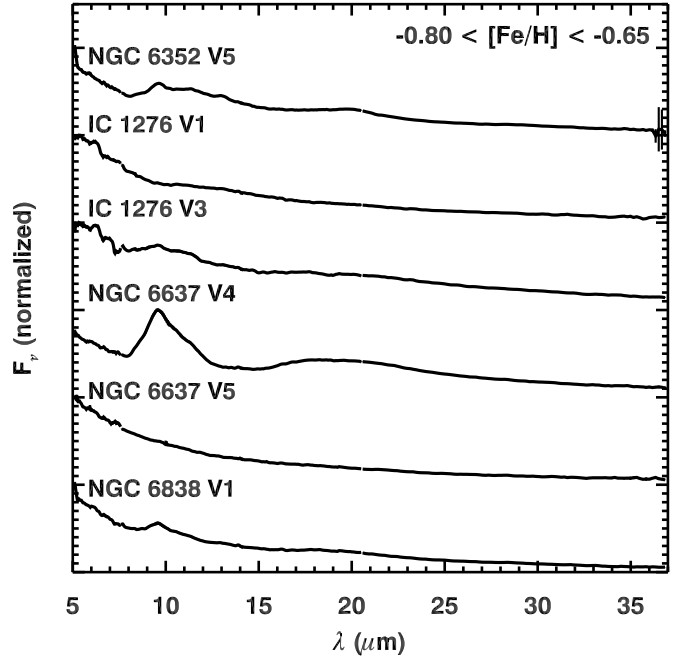}
\caption{IRS spectra of the six sources in our sample with
metallicities ([Fe/H]) from $-$0.65 to $-$0.80.}
\end{figure} \label{FigSp4}

\begin{figure} % Fig. 6
\includegraphics[width=3.5in]{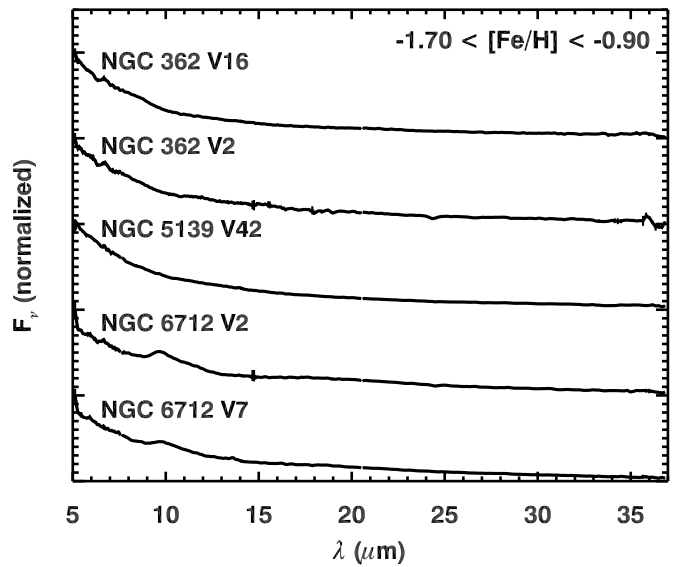}
\caption{IRS spectra of the five AGB variables in our sample
with metallicities ([Fe/H]) below $-$0.90.} \label{FigSp5}
\end{figure}

\begin{figure} % Fig. 7
\includegraphics[width=3.5in]{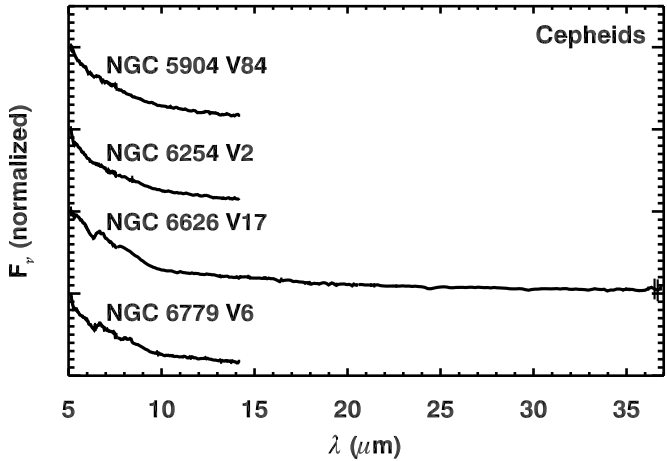}
\caption{IRS spectra of the four Cepheid variables in our 
sample.  All have metallicities ([Fe/H]) below $-$1.0.  Only 
one of the four has LL data.} \label{FigSp6}
\end{figure}

\begin{figure} % Fig. 8
\includegraphics[width=3.5in]{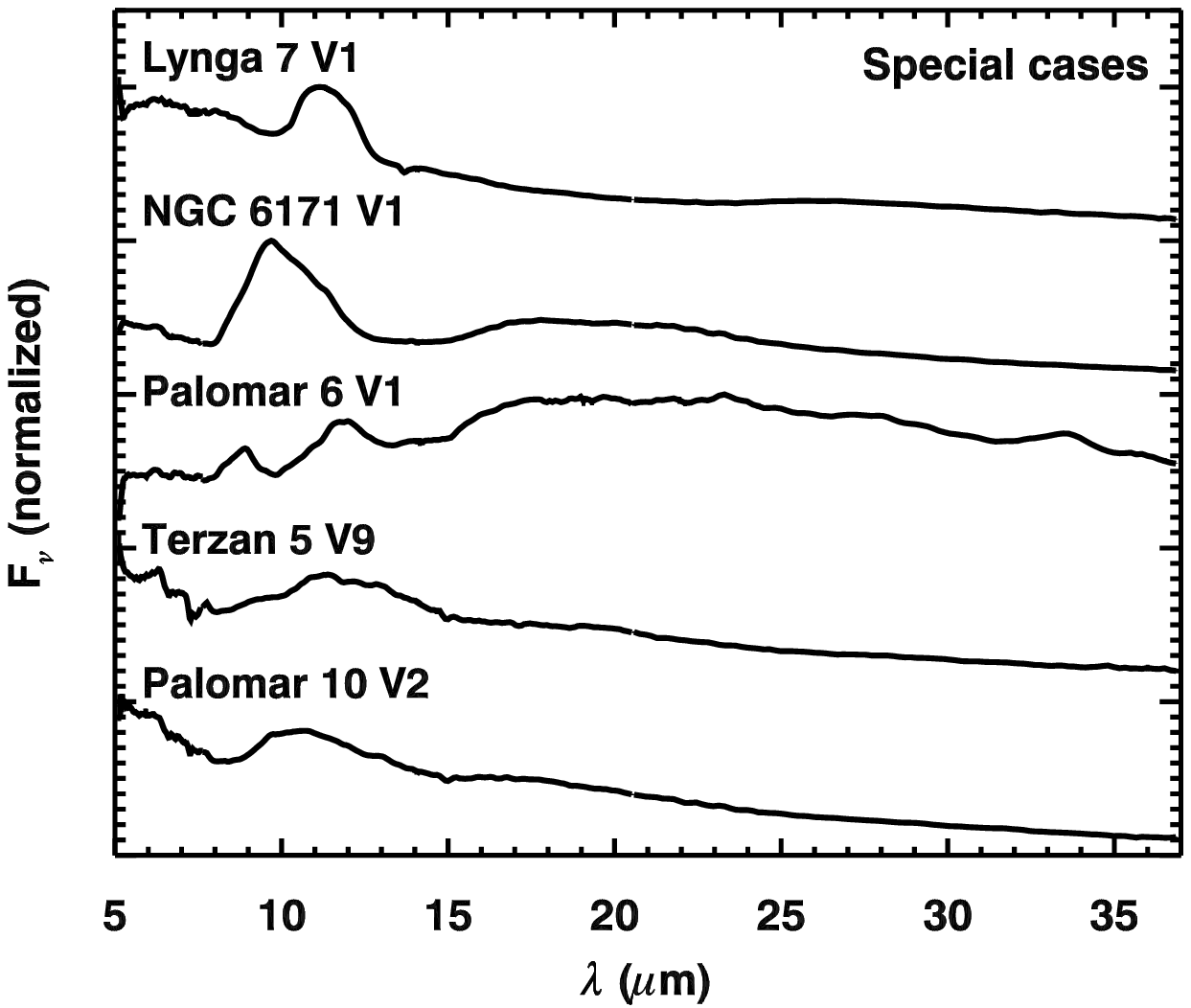}
\caption{IRS spectra of the five sources which cannot be
included in the metallicity-dependent analysis, as
explained in \S~\ref{SecSpecial}.}
\label{FigSp7}
\end{figure}

We observed the sample of variables in Table~\ref{TblVar} 
with the IRS on {\it Spitzer}, using the Short-Low (SL) 
module on all targets and the Long-Low (LL) module on all but 
the three faintest.  The data reduction began with the 
flatfielded images produced by the S18 version of the data 
pipeline from the {\it Spitzer} Science Center (SSC).  To 
subtract the background from these images, we generally used 
images with the source in the other aperture in SL (aperture 
differences) and images with the source in the other nod 
position in the same aperture in LL (nod differences).  In 
crowded fields, we occasionally had to resort to whichever 
background image resulted in the least confusion.  We then 
corrected the bad pixels in the differenced images using the 
{\rm imclean} IDL package\footnote{Available from the SSC as 
{\rm irsclean}.}.  Bad pixels included those flagged in the 
bit-mask images accompanying the data images and those 
flagged as rogues in the campaign rogue masks provided by the 
SSC.  We stacked the campaign rogue masks to produce 
super-rogue masks, considering a pixel to be bad if it had 
been identified as such in any two previous campaign rogue 
masks.

We used two methods to extract spectra from the corrected 
images.  The older method uses the {\rm profile}, {\rm 
ridge}, and {\rm extract} routines available in the SSC's 
{\it Spitzer} IRS Custom Extractor (SPICE).  This method 
produces the equivalent of a tapered-column extraction in 
SMART \citep{hig04}\footnote{SMART is the Spectroscopic
Modeling, Analysis, and Reduction Tool.}, summing the flux 
within an aperture which increases in width proportionally 
with wavelength.  It is the basis for the spectra in most of 
the IRS papers referenced in the introduction.  This globular 
cluster data-set is one of the first to make use of the 
optimal extraction algorithm available in the new release of 
SMART \citep{leb10}.  This algorithm fits a super-sampled 
point-spread function (PSF) to the data, improving the 
signal/noise (S/N) ratio of the data by a factor of 
$\sim$1.8, typically.  Optimal extraction also allows the 
extraction of sources with overlapping PSFs, which is 
particularly useful in the crowded regions typical for 
globular clusters.  

We used the optimal extraction for all of our data except for
the LL portion of three targets:  NGC~362~V16, Terzan~5~V5,
and NGC~6441~V2.  In these cases, the optimal algorithm was
unable to separate our intended target from adjacent sources,
and for these, we used the tapered-column extraction.

The calibration of SL spectra is based on observations of 
HR 6348 (K0 III).  The calibration of LL used HR 6348 along 
with the late K giants HD 166780 and HD 173511.  We extracted 
the spectra of these stars using the optimal and 
tapered-column algorithms to calibrate the two extraction 
methods independently.

The final step in the data reduction corrects for 
discontinuities between the orders and trims untrustworthy 
data from the ends of each order.  The correction for 
discontinuities applies scalar multiplicative adjustments to 
each order, shifting them upwards to the presumably 
best-centered segment.  The exception is the three targets 
blended with other sources in LL.  For these, we normalized 
to SL1 to minimize the impact on our measured bolometric 
magnitudes of the contamination from the additional sources 
in LL.

Figures~\ref{FigSp1} through \ref{FigSp7} present the 
resulting spectra for the targets in Table~\ref{TblVar}, 
organized by the metallicity of the clusters of which they 
are members (and ordered by the right ascension of the 
cluster).  Figure~\ref{FigSp6} includes the four Cepheids, 
three of which were observed with SL only, and 
Figure~\ref{FigSp7} includes those sources which are not 
oxygen-rich (Lyng{\aa}~7~V1), are members of clusters with 
highly uncertain metallicities (Palomar~6~V1), or whose 
membership or evolutionary status is uncertain 
(\S~\ref{SecMemSum}).

\subsection{Contemporaneous Photometry} % Sec. 2.4
\label{SecPhot}

\begin{deluxetable*}{lrrrrl} % Table 3
%\begin{deluxetable}{lrrrrl}
%\tabletypesize{\small}
\tablecolumns{6}
\tablewidth{0pt}
\tablecaption{Photometry from Siding Spring Observatory\label{TblSSO}}

\tablehead{
  \colhead{Target} & \colhead{J} & \colhead{H} & \colhead{K} & \colhead{L} & \colhead{JD}
}
\startdata
NGC 362 V16    &  9.384 $\pm$ 0.011 &  8.671 $\pm$ 0.007 & 8.469 $\pm$ 0.006 & 8.046 $\pm$ 0.036 & 2454376 \\
NGC 362 V2     &  9.781 $\pm$ 0.011 &  9.017 $\pm$ 0.007 & 8.829 $\pm$ 0.005 & 8.571 $\pm$ 0.056 & 2454376 \\
NGC 5139 V42   &  8.694 $\pm$ 0.011 &  7.908 $\pm$ 0.037 & 7.583 $\pm$ 0.032 & 6.974 $\pm$ 0.022 & 2454376 \\
NGC 5904 V84   &  9.931 $\pm$ 0.008 &  9.587 $\pm$ 0.009 & 9.525 $\pm$ 0.008 & 9.308 $\pm$ 0.115 & 2454376 \\
NGC 5927 V1    &  9.422 $\pm$ 0.013 &  8.379 $\pm$ 0.005 & 7.961 $\pm$ 0.004 & 7.387 $\pm$ 0.049 & 2454376 \\
NGC 5927 V3    &  8.309 $\pm$ 0.011 &  7.289 $\pm$ 0.006 & 6.869 $\pm$ 0.006 & 6.238 $\pm$ 0.018 & 2454376 \\
Lyng{\aa} 7 V1 & 11.653 $\pm$ 0.021 &  9.308 $\pm$ 0.019 & 7.317 $\pm$ 0.009 & 4.896 $\pm$ 0.017 & 2454376 \\
NGC 6171 V1    &  5.842 $\pm$ 0.007 &  5.069 $\pm$ 0.040 & 4.623 $\pm$ 0.010 & 3.772 $\pm$ 0.018 & 2454376 \\
NGC 6254 V2    &  9.749 $\pm$ 0.018 &  9.322 $\pm$ 0.017 & 9.231 $\pm$ 0.017 & 9.004 $\pm$ 0.099 & 2454376 \\
NGC 6356 V1    &  9.986 $\pm$ 0.015 &  9.136 $\pm$ 0.010 & 8.674 $\pm$ 0.006 & 8.063 $\pm$ 0.038 & 2454375 \\
NGC 6356 V3    & 10.490 $\pm$ 0.029 &  9.619 $\pm$ 0.015 & 9.151 $\pm$ 0.008 & 8.649 $\pm$ 0.067 & 2454375 \\
NGC 6356 V4    & 10.352 $\pm$ 0.022 &  9.458 $\pm$ 0.014 & 9.118 $\pm$ 0.007 & 8.683 $\pm$ 0.079 & 2454375 \\
NGC 6356 V5    & 10.030 $\pm$ 0.019 &  9.113 $\pm$ 0.016 & 8.777 $\pm$ 0.005 & 8.231 $\pm$ 0.048 & 2454375 \\
NGC 6388 V3    & 10.303 $\pm$ 0.025 &  9.203 $\pm$ 0.026 & 8.957 $\pm$ 0.013 & 8.602 $\pm$ 0.064 & 2454376 \\
NGC 6388 V4    &  9.781 $\pm$ 0.007 &  8.926 $\pm$ 0.005 & 8.563 $\pm$ 0.017 & 7.925 $\pm$ 0.039 & 2454376 \\
Palomar 6 V1   & 14.010 $\pm$ 0.078 & 10.508 $\pm$ 0.029 & 8.036 $\pm$ 0.019 & 5.471 $\pm$ 0.030 & 2454376 \\
Terzan 5 V2    &  9.980 $\pm$ 0.013 &  8.515 $\pm$ 0.008 & 7.802 $\pm$ 0.006 & 6.947 $\pm$ 0.021 & 2454376 \\
Terzan 5 V5    &  9.413 $\pm$ 0.011 &  7.496 $\pm$ 0.007 & 6.511 $\pm$ 0.005 & 5.472 $\pm$ 0.016 & 2454376 \\
Terzan 5 V6    & 10.323 $\pm$ 0.028 &  8.609 $\pm$ 0.013 & 7.724 $\pm$ 0.017 & 6.712 $\pm$ 0.021 & 2454376 \\
Terzan 5 V7    &  9.369 $\pm$ 0.007 &  7.763 $\pm$ 0.007 & 6.838 $\pm$ 0.004 & 5.724 $\pm$ 0.016 & 2454376 \\
Terzan 5 V8    &  9.764 $\pm$ 0.032 &  8.257 $\pm$ 0.029 & 7.467 $\pm$ 0.008 & 6.589 $\pm$ 0.018 & 2454376 \\
Terzan 5 V9    & 11.307 $\pm$ 0.013 &  9.301 $\pm$ 0.009 & 8.173 $\pm$ 0.007 & 7.006 $\pm$ 0.029 & 2454376 \\
NGC 6553 V4    &  7.905 $\pm$ 0.020 &  6.735 $\pm$ 0.014 & 6.298 $\pm$ 0.015 & 5.787 $\pm$ 0.038 & 2454375 \\
%NGC 6553 V5   &  7.752 $\pm$ 0.006 &  6.620 $\pm$ 0.008 & 6.221 $\pm$ 0.013 & 5.798 $\pm$ 0.016 & 2454376 \\
IC 1276 V1     &  8.604 $\pm$ 0.013 &  7.417 $\pm$ 0.008 & 6.882 $\pm$ 0.005 & 6.283 $\pm$ 0.017 & 2454375 \\
IC 1276 V3     &  8.926 $\pm$ 0.013 &  7.786 $\pm$ 0.009 & 7.015 $\pm$ 0.010 & 6.083 $\pm$ 0.048 & 2454375 \\
Terzan 12 V1   &  8.677 $\pm$ 0.021 &  7.096 $\pm$ 0.024 & 6.207 $\pm$ 0.015 & 5.164 $\pm$ 0.017 & 2454375 \\
NGC 6626 V17   &  9.133 $\pm$ 0.024 &  8.585 $\pm$ 0.018 & 8.394 $\pm$ 0.033 & 8.271 $\pm$ 0.044 & 2454375 \\
NGC 6760 V3    &  8.918 $\pm$ 0.006 &  7.898 $\pm$ 0.009 & 7.457 $\pm$ 0.004 & 6.869 $\pm$ 0.039 & 2454376 \\
NGC 6760 V4    &  9.779 $\pm$ 0.008 &  8.739 $\pm$ 0.006 & 8.142 $\pm$ 0.004 & 7.356 $\pm$ 0.023 & 2454376 \\
Palomar 10 V2  &  7.227 $\pm$ 0.006 &  5.984 $\pm$ 0.008 & 5.292 $\pm$ 0.005 & 4.439 $\pm$ 0.021 & 2454376 \\
NGC 6838 V1    &  7.449 $\pm$ 0.006 &  6.597 $\pm$ 0.005 & 6.283 $\pm$ 0.004 & 5.830 $\pm$ 0.016 & 2454376 \\
\enddata
\end{deluxetable*}
%\end{deluxetable}

Table~\ref{TblSSO} presents photometry of 31 of our 39 
targets obtained from the 2.3-m telescope at Siding Spring 
Observatory (SSO) during the period when the {\it Spitzer} 
spectra were obtained.  All observations were made using the 
Cryogenic Array Spectrometer/Imager \citep[CASPIR;][]{caspir} 
with the following filters:  J (effective wavelength 
1.24~\mum), H (1.68~\mum), K (2.22~\mum), and narrow-band L 
(3.59~\mum).  Calibrations were based on observations of 
standard stars from the lists of \cite{mcg94}, and the data 
were processed using standard tools available with IRAF (the 
Image Reduction and Analysis Facility).

\section{Membership} % Sec. 3
\label{SecMem}

The principal objective of this study is to investigate how 
the quantity and composition of the dust produced by AGB 
stars depend on its initial metallicity.  We assume that 
the metallicity of a star is simply the metallicity of its
cluster, making it important to identify non-members which
are in the foreground or the background of a cluster.  
Our primary tools for testing membership are whether the 
relation between period and absolute K magnitude and the 
bolometric magnitude of the stars produces a distance
consistent with membership, and if that distance results in
a bolometric magnitude consistent with the masses of stars
as old as those in our sample.

\subsection{The Period-K Relation} % Sec. 3.1
\label{SecPK}

\begin{figure} % Fig. 9
\includegraphics[width=3.5in]{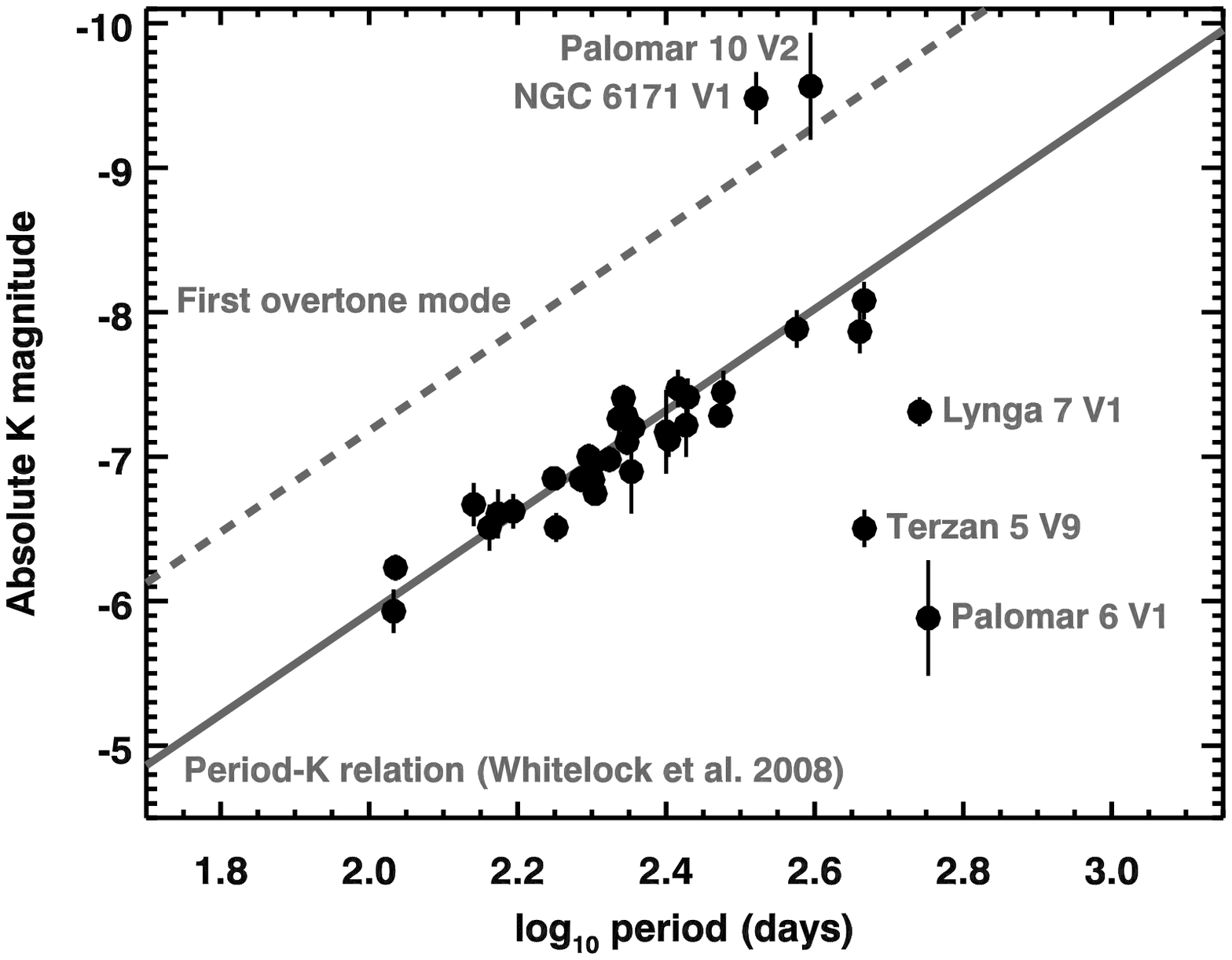}
\caption{The periods and absolute K magnitudes of the 
variables in our globular cluster sample (excluding Cepheids).  
Absolute magnitudes are determined from the distance moduli 
for the clusters reported in Table~\ref{TblClust}.  The 
solid line is the nominal period-K relation defined by 
\cite{whi08}.  The dashed line represents the shift in the 
relation expected for stars pulsating in the first overtone 
mode.  Sources which do not follow the period-K relation are 
labelled (see \S~\ref{SecMemSum}).} 
\label{FigPK}
\end{figure}

We estimated the distance to each variable by applying the 
period-K magnitude relation of \cite{whi08}.
In general, $M_K$ = $\rho$ (log $P$ $-$ 2.38) + $\delta$.
\cite{whi08} found that $\rho$ = $-$3.51 in Mira variables in
the LMC and the Galaxy, but $\delta$
varied from $-$7.15 in the LMC to $-$7.25 in the Galaxy.
\cite{whi08} assumed that the distance modulus to the LMC
was 18.39.  The value for $\delta$ would be $-$7.25 for both 
the Galactic and LMC samples if the distance modulus to the
LMC were 18.49, which compares favorably to recent
measurements.  \cite{alv04} estimated the distance modulus to
the LMC to be 18.50$\pm$0.02, based on a review of 14 recent
determinations, and \cite{kw06} found a value of 
18.54$\pm$0.02, based on an analysis of bump Cepheids.  Thus,
we assume that $\delta$=$-$7.25 for all sources in the 
following analysis.

We dereddened the K magnitude using the estimates for 
$E(B-V)$ given in Table~\ref{TblClust} and the interstellar 
extinction measurements by \cite{rl85}, who find that 
$R = A_V / E(B-V) = 3.09 \pm 0.03$ and 
$A_K/A_V = 0.112$\footnote{We will also use the ratios 
$A_J/A_V = 0.282$, $A_H/A_V = 0.175$, and $A_L/A_V = 0.058$ 
elsewhere in this paper.}  We excluded the four Cepheid 
variables in our sample from this analysis, which accounts 
for all of the periods less than 50 days.  

The period-K relation produces distance moduli for IC~1276~V1 
and V3 of 13.51 and 13.81, respectively, while \cite{har03} 
reports 13.66 and \cite{bar98b} report 13.01.  The latter is 
inconsistent with our estimates, and we adopt the former.
Similarly, the period-K relation gives a distance modulus
for Terzan~12~V1 of 13.75, compared to 13.38 \citep{har03}
and 12.66 \citep{ort98}.  This source has a period of 458
days, and we might expect circumstellar extinction at K of
a few tenths of a magnitude.  Thus we drop 12.66 as
inconsistent with the period of the star and adopt 13.38
as the distance modulus.  

Five of the six variables in Terzan~5 have an average 
distance modulus of 14.11 ($\pm$0.13), compared to values of 
15.05 \citep{har03}, 13.87 \citep{val07}, and 13.70 
\citep{ort07}.  Terzan~5~V9 is heavily reddened at K, giving 
an apparent distance modulus of 15.86, widely at variance 
with the others.  Differential reddening in front of 
Terzan~5 would have to produce 15.6 additional magnitudes of 
visual extinction and is an unlikely explanation.  
\cite{fer09} have identified a younger and more metal-rich 
population in Terzan~5 (age$\sim$ 6 Gyr and 
[Fe/H]$\sim$+0.3).  They present isochrones showing that
the younger population should have roughly one extra 
magnitude of reddening in V$-$K, but their isochrones are 
based on models which do not extend to the tip of the AGB.
We cannot rule out the possibility that Terzan~5~V9 is a
member of this younger population, but it clearly differs
from the remaining sources in the cluster and to be 
cautious, we will treat it as a non-member.  The distance 
moduli of the remaining five variables are mutually 
consistent with each other and nearly two standard 
deviations from the closest published estimate, 13.87.  For 
the remainder of this paper, we will adopt our own mean 
distance modulus for Terzan~5:  14.11.

Figure~\ref{FigPK} presents the period-K relation for our 
sample of globular cluster variables (excluding the 
Cepheids).  The absolute K magnitudes are based on the 
distance moduli presented in Table~\ref{TblClust}.  The
labelled sources in the figure require special consideration,
which in some cases has led to their exclusion as members.  
In Figure~\ref{FigPK}, though, their absolute magnitude is 
determined as though they were members.  The figure also 
shows the period-K relation for fundamental-mode pulsators 
and for overtone pulsators, assuming that their periods are 
2.3 times smaller.

Two sources appear to be too bright for the period-K relation: 
NGC~6171~V1 and Palomar~10~V2.  While it is possible that they 
are overtone pulsators, their bolometric magnitudes are 
inconsistent with membership, as explained in the next 
section.  Assuming that both are fundamental-mode pulsators, 
NGC~6171~V1 has a distance modulus of 12.15, compared to 
13.89 for the cluster, putting it 3.3 kpc in front of the 
cluster and 1.1 kpc above the Galactic plane.  Similarly, 
Palomar~10~V2 has a distance modulus of 12.30 vs.\ 13.86 
for the cluster, which means it is 3.0 kpc closer and about 
0.14 kpc above the Galactic plane.

The three sources below the period-K relation present more
complex cases.  We will examine them after determining
bolometric magnitudes (next section).

Figure~\ref{FigPK} also shows that all of the confirmed 
members are pulsating in the fundamental mode, even though 
several sources are classified as semi-regular variables, or 
possible semi-regular variables, in Table~\ref{TblVar}.  
These results are consistent if they are SRa variables.

\subsection{Bolometric Magnitudes} % Sec. 3.2
\label{SecBol}

\begin{figure} % Fig. 10
\includegraphics[width=3.5in]{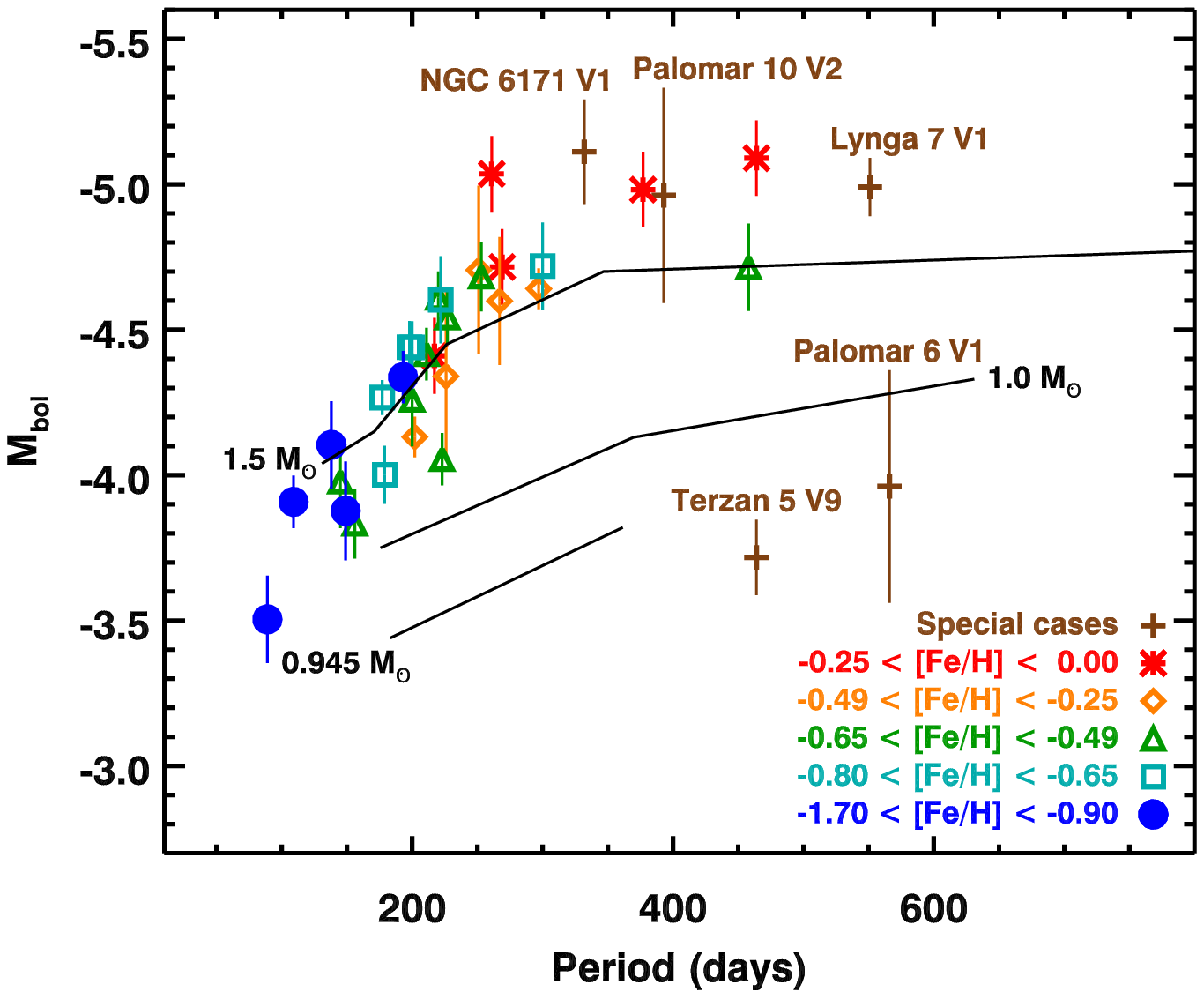}
\caption{The bolometric magnitudes and pulsation periods of
the globular sample, compared to models of AGB stars at
0.945, 1.0, and 1.5 M$_{\sun}$ \citep[black lines;][]{vw93}.
The symbols are color-coded to run from the most metal-rich 
globular cluster (red) to the most metal-poor (blue), with 
special cases in brown.} 
\label{FigPM}
\end{figure}

\begin{deluxetable*}{lrrrllll} % Table 4
%\begin{deluxetable}{lrrlllll}
%\rotate
%\tabletypesize{\small}
\tablecolumns{8}
\tablewidth{0pt}
\tablecaption{Spectroscopic Properties\label{TblSp}}
\tablehead{
  \colhead{ } & \colhead{ } & \colhead{ } & \colhead{ } & \colhead{log \mdot} &
  \colhead{Corrected} & \colhead{IR Spec.} & \colhead{ } \\
  \colhead{Target} & \colhead{$M_{bol}$} & \colhead{[7]$-$[15]} & 
  \colhead{DEC\tablenotemark{a}} & \colhead{(M$_{\sun}$ /yr)\tablenotemark{b}} &
  \colhead{$F_{11}/F_{12}$} & \colhead{Class\tablenotemark{c}} & \colhead{Notes}\\
}
\startdata
NGC 362 V2     & $-$3.50 & 0.46 $\pm$ 0.02 &    0.09 $\pm$ 0.01 & $-$7.90 $\pm$ 0.06 & \nodata         & 1.N      \\
NGC 362 V16    & $-$4.10 & 0.25 $\pm$ 0.02 &    0.02 $\pm$ 0.00 & $-$8.54 $\pm$ 0.44 & \nodata         & 1.N      \\
NGC 5139 V42   & $-$3.88 & 0.41 $\pm$ 0.01 &    0.07 $\pm$ 0.00 & $-$8.02 $\pm$ 0.11 & \nodata         & 1.N      \\
NGC 5904 V84   & $-$3.05 & \nodata         & $-$0.06 $\pm$ 0.03 & \nodata            & \nodata         & 1.N     & d, e \\
NGC 5927 V1    & $-$4.13 & 1.19 $\pm$ 0.02 &    0.67 $\pm$ 0.01 & $-$6.65 $\pm$ 0.11 & 1.22 $\pm$ 0.01 & 2.SX4t   \\
NGC 5927 V3    & $-$4.64 & 1.35 $\pm$ 0.00 &    1.48 $\pm$ 0.00 & $-$6.27 $\pm$ 0.03 & 1.57 $\pm$ 0.05 & 2.SE8t   \\
Lyng{\aa} 7 V1 & $-$4.99 & \nodata         &    0.58 $\pm$ 0.01 & $-$6.61 $\pm$ 0.01 & \nodata         & 2.CE     \\
NGC 6171 V1    & $-$5.11 & 1.53 $\pm$ 0.01 &    1.99 $\pm$ 0.01 & $-$6.02 $\pm$ 0.07 & 1.68 $\pm$ 0.06 & 2.SE8   & f \\
NGC 6254 V2    & $-$1.99 & \nodata         & $-$0.02 $\pm$ 0.01 & \nodata            & \nodata         & 1.N     & d, e \\
NGC 6352 V5    & $-$4.27 & 1.05 $\pm$ 0.02 &    0.59 $\pm$ 0.01 & $-$6.81 $\pm$ 0.01 & 1.16 $\pm$ 0.03 & 2.SX4t   \\
NGC 6356 V1    & $-$4.54 & 1.47 $\pm$ 0.01 &    2.08 $\pm$ 0.01 & $-$6.06 $\pm$ 0.03 & 1.62 $\pm$ 0.06 & 2.SE8f   \\
NGC 6356 V3    & $-$4.06 & 0.58 $\pm$ 0.02 &    0.17 $\pm$ 0.01 & $-$7.66 $\pm$ 0.02 & 0.83 $\pm$ 0.04 & 2.SE1    \\
NGC 6356 V4    & $-$4.42 & 0.61 $\pm$ 0.02 &    0.14 $\pm$ 0.01 & $-$7.57 $\pm$ 0.03 & 1.03 $\pm$ 0.05 & 2.SE2 t: \\
NGC 6356 V5    & $-$4.61 & 0.67 $\pm$ 0.02 &    0.25 $\pm$ 0.00 & $-$7.40 $\pm$ 0.12 & 1.37 $\pm$ 0.04 & 2.SE6 t: \\
NGC 6388 V3    & $-$3.83 & 0.13 $\pm$ 0.02 & $-$0.04 $\pm$ 0.01 & $-$8.72 $\pm$ 0.40 & \nodata         & 1.N      \\
NGC 6388 V4    & $-$4.68 & 0.91 $\pm$ 0.01 &    0.68 $\pm$ 0.04 & $-$6.89 $\pm$ 0.25 & 1.73 $\pm$ 0.07 & 2.SE8    \\
Palomar 6 V1   & $-$3.96 & 2.02 $\pm$ 0.01 &    1.27 $\pm$ 0.01 & $-$5.11 $\pm$ 0.43 & \nodata         & 3.SBxf  & g \\
Terzan 5 V2    & $-$4.41 & 0.62 $\pm$ 0.01 &    0.11 $\pm$ 0.01 & $-$7.70 $\pm$ 0.17 & 0.79 $\pm$ 0.07 & 2.SE1 t: \\
Terzan 5 V5    & $-$5.09 & 1.38 $\pm$ 0.02 &    0.91 $\pm$ 0.02 & $-$6.39 $\pm$ 0.22 & 1.23 $\pm$ 0.02 & 2.SE4    \\
Terzan 5 V6    & $-$4.72 & 1.60 $\pm$ 0.01 &    1.41 $\pm$ 0.01 & $-$6.06 $\pm$ 0.30 & 1.17 $\pm$ 0.01 & 2.SX4t   \\
Terzan 5 V7    & $-$4.98 & 1.61 $\pm$ 0.01 &    1.71 $\pm$ 0.00 & $-$6.00 $\pm$ 0.23 & 1.32 $\pm$ 0.03 & 2.SE5    \\ % SB?
Terzan 5 V8    & $-$5.04 & 0.78 $\pm$ 0.01 &    0.15 $\pm$ 0.01 & $-$7.46 $\pm$ 0.24 & 0.21 $\pm$ 0.11 & 2.SE1 t: \\
Terzan 5 V9    & $-$3.72 & 1.46 $\pm$ 0.03 &    0.88 $\pm$ 0.01 & $-$6.33 $\pm$ 0.33 & 0.95 $\pm$ 0.03 & 2.SY2t  & g \\
NGC 6441 V1    & $-$4.26 & 0.71 $\pm$ 0.01 &    0.20 $\pm$ 0.01 & $-$7.44 $\pm$ 0.03 & 0.84 $\pm$ 0.04 & 2.SE1t   \\
NGC 6441 V2    & $-$3.98 & 0.52 $\pm$ 0.03 &    0.12 $\pm$ 0.02 & $-$7.76 $\pm$ 0.01 & 0.99 $\pm$ 0.14 & 2.SE2    \\
NGC 6553 V4    & $-$4.60 & 0.90 $\pm$ 0.01 &    0.32 $\pm$ 0.01 & $-$8.60 $\pm$ 0.57 & 0.94 $\pm$ 0.02 & 2.SY1t   \\
IC 1276 V1     & $-$4.60 & 0.71 $\pm$ 0.01 &    0.13 $\pm$ 0.01 & $-$7.57 $\pm$ 0.21 & 0.68 $\pm$ 0.08 & 2.SE1 t: \\
IC 1276 V3     & $-$4.72 & 1.02 $\pm$ 0.02 &    0.52 $\pm$ 0.01 & $-$6.88 $\pm$ 0.01 & 1.22 $\pm$ 0.01 & 2.SE4    \\
Terzan 12 V1   & $-$4.72 & 1.48 $\pm$ 0.02 &    1.06 $\pm$ 0.01 & $-$6.26 $\pm$ 0.27 & 1.13 $\pm$ 0.02 & 2.SY3    \\
NGC 6626 V17   & $-$3.42 & 0.27 $\pm$ 0.02 & $-$0.08 $\pm$ 0.01 & $-$8.60 $\pm$ 0.57 & \nodata         & 1.N     & d \\
NGC 6637 V4    & $-$4.44 & 1.06 $\pm$ 0.01 &    1.12 $\pm$ 0.02 & $-$6.61 $\pm$ 0.27 & 1.70 $\pm$ 0.08 & 2.SE8    \\
NGC 6637 V5    & $-$4.44 & 0.56 $\pm$ 0.01 &    0.15 $\pm$ 0.01 & $-$7.66 $\pm$ 0.03 & 1.17 $\pm$ 0.04 & 2.SE4    \\
NGC 6712 V2    & $-$3.91 & 0.56 $\pm$ 0.02 &    0.27 $\pm$ 0.01 & $-$7.48 $\pm$ 0.29 & 1.62 $\pm$ 0.08 & 2.SE8    \\
NGC 6712 V7    & $-$4.34 & 0.52 $\pm$ 0.01 &    0.21 $\pm$ 0.01 & $-$7.59 $\pm$ 0.23 & 1.51 $\pm$ 0.05 & 2.SE7    \\
NGC 6760 V3    & $-$4.71 & 0.78 $\pm$ 0.01 &    0.51 $\pm$ 0.02 & $-$7.09 $\pm$ 0.28 & 1.72 $\pm$ 0.07 & 2.SE8    \\
NGC 6760 V4    & $-$4.34 & 0.96 $\pm$ 0.01 &    0.60 $\pm$ 0.01 & $-$6.88 $\pm$ 0.13 & 1.33 $\pm$ 0.02 & 2.SE5    \\
NGC 6779 V6    & $-$2.96 & \nodata         & $-$0.06 $\pm$ 0.01 & \nodata            & \nodata         & 1.N     & d, e \\
Palomar 10 V2  & $-$4.96 & 1.21 $\pm$ 0.02 &    0.68 $\pm$ 0.00 & $-$6.63 $\pm$ 0.13 & 1.14 $\pm$ 0.02 & 2.SE3t  & h \\ % SB3t
NGC 6838 V1    & $-$4.00 & 0.77 $\pm$ 0.01 &    0.36 $\pm$ 0.01 & $-$7.21 $\pm$ 0.15 & 1.42 $\pm$ 0.04 & 2.SE6 t:  \\
\enddata
\tablenotetext{a}{Dust emission contrast; see \S 4.1 for an explanation and
   a discussion of systematic errors.}
\tablenotetext{b}{Subtract 2.30 to convert to log$_{10}$ of the dust-production 
rate (see \S 4.2).}
\tablenotetext{c}{These classifications are defined in \S 4.1.}
\tablenotetext{d}{Cepheid variable.}
\tablenotetext{e}{SL data only.}
\tablenotetext{f}{Assumed $m-M$ = 12.15.}
\tablenotetext{g}{Membership uncertain.}
\tablenotetext{h}{Assumed $m-M$ = 12.30.}
\end{deluxetable*}
%\end{deluxetable}

Table~\ref{TblSp} includes bolometric magnitudes for each
star in our sample.  Basically, we integrated the photometry
from SSO and the spectroscopic data from {\it Spitzer} and 
corrected for the distances in Table~\ref{TblClust} (except 
for the two most likely non-members, NGC~6171~V1 and
Palomar~10~V2, where we used the distance moduli determined 
in the preceding section).  For the eight stars not observed 
from SSO, we substituted the mean magnitudes in 
Table~\ref{TblVar}.  Most of the SSO photometry was taken 
within one or two days of the {\it Spitzer} observations, but 
when this was not the case (and for the non-SSO photometry), 
we phase-corrected the data to match the {\it Spitzer} epoch, 
using the amplitudes from our analysis of the IRSF/SIRIUS
monitoring data.  The monitoring data did not include 
observations at L, and we assumed that $\Delta$L$\sim 
\Delta$K, as confirmed by the mean amplitudes for oxygen-rich 
LPVs published by \cite{smi03}.  We also corrected the 
photometry for interstellar extinction, using the E(B$-$V) 
excesses in Table~\ref{TblClust} and the insterstellar 
reddening of \cite{rl85}.  To integrate the flux outside of 
the observed wavelengths, we extended a 3600~K blackbody to 
the blue and a Rayleigh-Jeans tail to the red.  Finally, to 
correct the bolometric magnitudes for pulsation, we assumed 
that the K-band amplitude approximated the bolometric 
amplitude and used the phase to correct to the mean.

Figure~\ref{FigPM} plots the bolometric magnitudes of our 
sample against their pulsation periods, and it includes
(simplified) evolutionary tracks taken from \citet[][Fig.\ 
20]{vw93}.  We will return to this diagram in 
\S~\ref{SecEvo}.  Here, the diagram can help us assess 
cluster membership.  Except for two sources in the lower
right (Terzan~5~V9 and Palomar~6~V1), the entire sample
appears to follow a single sequence from the lower left
to the upper right.  Terzan~12~V1 is the orange diamond at
the lower edge of the sequence with a period of 458 days.
It is close enough to the sequence defined by the 
remaining sources that it validates our choice of
distance modulus.  The next section examines the three 
sources that appear below the period-K relation 
(Figure~\ref{FigPK}) in turn.

\subsection{Special Cases} % Sec. 3.3
\label{SecSpecial}

The bolometric magnitudes of NGC~6171~V1 and Palomar~10~V2 in 
Table~\ref{TblSp} are based on the presumption that they
are foreground objects (\S~\ref{SecPK}).  If they were 
actually overtone pulsators and cluster members, their 
bolometric magnitudes would be $-$6.85 and $-$6.52, 
respectively.  According to the evolutionary models of 
\cite{vw93}, these magnitudes would correspond to initial 
masses of $\sim$5~M$_{\sun}$, much too massive for the age of 
either cluster.

Terzan~5~V9 is about 1.7 magnitudes too faint at K if it were
at the distance of the cluster.  Its J$-$K and H$-$K colors, 
corrected for interstellar extinction, are consistent with 
1.2--1.3 magnitudes of circumstellar extinction at K, and its 
K$-$L color suggests A$_K$$\sim$1.6.  These color-based 
estimates assume that the circumstellar dust extinction 
follows the interstellar relationships of \cite{rl85}, and 
they are roughly consistent with the observed extinction.  
However, this source sits in the lower right in 
Figure~\ref{FigPM} with P=464 days.  If it really is a member 
of Terzan~5, its bolometric magnitude is inconsistent with 
its pulsation period.  Perhaps it is a background object, or 
perhaps it is a binary or interacting system.  Whatever it 
is, it is not a normal AGB star, and we consequently do not 
include it when analyzing the other AGB stars in Terzan~5.

Palomar~6~V1 is too faint by 2.6 magnitudes at K, but its 
J$-$K, H$-$K, and K$-$L colors correspond to circumstellar 
extinction at K of 3.5--4.9 magnitudes.  These estimates
would put the source $\sim$1--2 magnitudes of distance
modulus {\it in front of} Palomar~6.  In Figure~\ref{FigPM}, 
Palomar~6~V1 is the source below the others in the lower 
right with P=566 days.  If it were in the foreground of 
Palomar~6, correcting for its distance would push it even 
further down in Figure~\ref{FigPM}.  Here again, the star 
cannot be a normal AGB star and a cluster member.  As 
already noted (\S \ref{SecClust}), even if it were,
the uncertain metallicity of the cluster 
prevents us from assigning it to one of our metallicity bins.

The spectrum of Lyng{\aa}~7~V1 (Figure~\ref{FigSp7}) is 
unambiguously that of a carbon star, with strong dust 
emission from SiC at $\sim$11.5~\mum\ and MgS dust at 
$\sim$26~\mum, as well as absorption from acetylene gas at 
7.5 and 13.7~\mum.  Its carbon-rich status may be consistent
with the circumstellar extinction at K apparent in 
Figure~\ref{FigPK}, but it would be helpful to independently
verify its distance.  

We have two means of doing so.  
\cite{slo08} calibrated a relationship of M$_K$ vs.\ J$-$K 
color for carbon stars in the SMC, using 2MASS photometry.
Correcting the mean magnitudes in Table~\ref{TblVar} for 
interstellar extinction gives J$-$K = 3.71, which leads to 
a distance modulus of 14.71.  \cite{slo08} also found and
calibrated a color-magnitude relation for the narrow 6.4- 
and 9.3-\mum\ filters defined to analyze earlier samples of 
Magellanic carbon stars.  Lyng{\aa}~7~V1 has a [6.4]$-$[9.3] 
color of 0.55, which implies M$_{9.3}$=$-$11.59.  Since 
[9.3]=2.67, the distance modulus is 14.26.  The average from 
these two methods is 14.20 $\pm$ 0.08, which compares 
favorably to the nominal distance modulus of Lyng{\aa}~7, 
14.31\footnote{Using an updated calibration of M$_K$ vs.\ 
J$-$K for the more metal-rich LMC \citep{lag10} gives $m-M$ 
= 14.71 and a mean $m-M$ = 14.49 $\pm$ 0.32, which is still 
close to the nominal distance.}.

In Figure~\ref{FigPM}, Lyng{\aa}~7~V1 is the upper right datum.
Its position is consistent with the evolutionary sequence
defined by the larger sample, adding some confidence that
it is a cluster member.  Nonetheless, its carbon-rich nature 
prevents any comparison with the oxygen-rich sample, and it 
remains excluded from most of the analysis in this paper.

% \cite{mat05} have detected SiO masers for four of our
% suspected non-members, and their radial velocities differ
% from the cluster by amounts ranging from 8 to 37 km/s.  While
% any of these radial velocities could indicate that the star
% was gravitationally bound to the cluster, a wide range of
% radial velocities  ...
% NGC 6171 V1 -28 km/s vs -20 km/s  Delta of 8 km/s
% Palomar 10 V2 14 km/s vs. -13 km/s  Delta of 27 km/s
% Terzan 12 V1 77 km/s vs. 106 km/s Delta of 29 km/s
% Palomar 6 V1 157 km/s vs. 194 km/s Delta of 37 km/s

\subsection{Membership Summary} % Sec. 3.4
\label{SecMemSum}

\begin{figure} % Fig. 11
\includegraphics[width=3.5in]{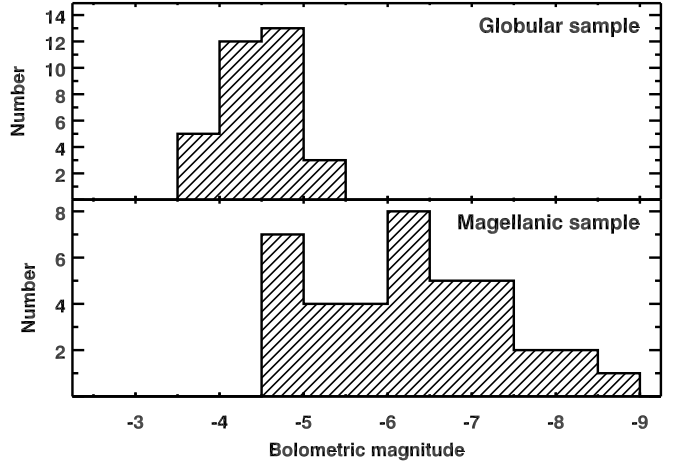}
\caption{A comparison of the bolometric magnitudes of
our globular sample of AGB stars and the the Magellanic 
sample of \cite{slo08}.  The histogram of the globular 
sample excludes the four Cepheid variables (all with
bolometric magnitudes between $-$1.6 and $-$2.7) and the
four sources whose cluster membership is in doubt.}
\label{FigHist}
\end{figure}

To summarize, we exclude NGC~6171~V1 and Palomar~10~V2 from
the sample because if they were cluster members, they would
be too bright for old AGB stars.  We exclude Terzan~5~V9 and
Palomar~6~V1 from further consideration because if they are
cluster members, they are too faint for their pulsation
periods compared to the rest of the sample.  Finally, while
Lyng{\aa}~7~V1 is a cluster member, it is a carbon star and 
must be treated separately from the rest of the sample.

Figure~\ref{FigHist} compares the bolometric magnitudes of 
the 31 LPVs in our sample which are confirmed as members of
globular clusters and normal AGB stars to the oxygen-rich
Magellanic sample considered by \cite{slo08}.  The
Magellanic sample spans a narrower range of metallicity 
($-$0.6 $\lesssim$ [Fe/H] $\lesssim$ $-$0.3).  It is readily 
apparent that the samples represent very different sources.  
The Magellanic sample consists primarily of supergiants and 
brighter AGB stars, while the globular sample contains only 
of fainter and lower-mass AGB stars.  This difference arises 
primarily from the greater distance to the Magellanic Clouds 
and the resulting bias toward more luminous sources.

\section{Spectral Analysis} % Sec. 4
\label{SecAnal}

\subsection{Infrared Spectral Classification} % Sec. 4.1

\begin{figure} % Fig. 12
\includegraphics[width=3.5in]{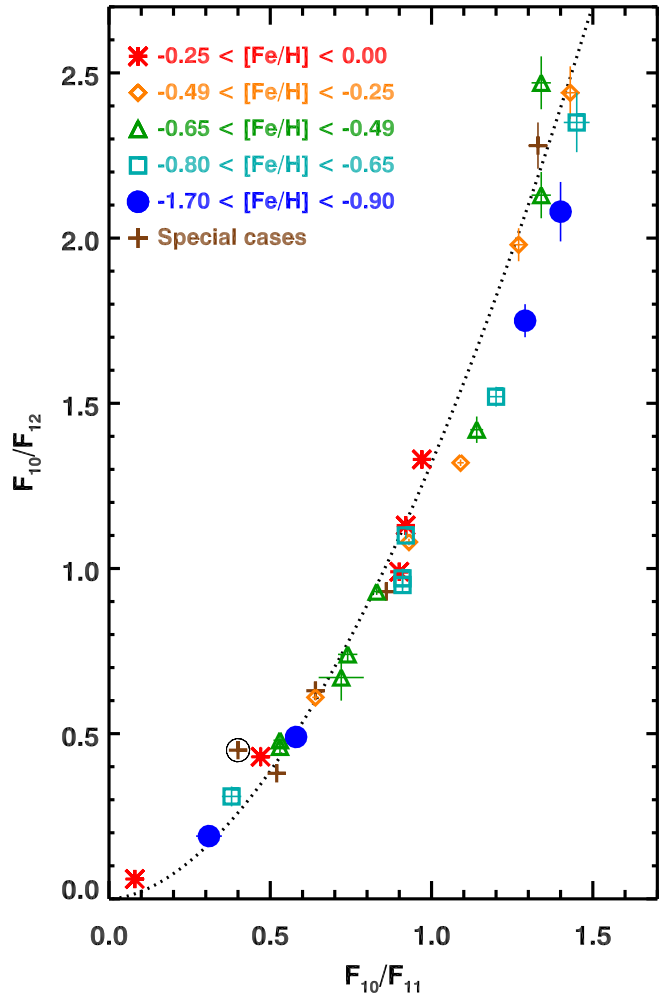}
\caption{The continuum-subtracted flux ratios of the globular 
sample compared to the silicate dust sequence (dashed line).  
The symbols are color-coded as in Fig.~\ref{FigPM}.
The circled brown cross in the lower left is the 
carbon star Lyng{\aa}~7~V1.  The four points to the right
of the silicate dust sequence (with 1.3 $<$ 
$F_{10}/F_{12}$ $<$ 1.8) are, from top to bottom,
NGC~6712~V7, NGC~6838~V1, NGC~6356~V5, and NGC~6760~V4,
and their deviation from the silicate dust sequence may
result from as-yet-unidentified impurities in the dust.}
\label{FigSDS}
\end{figure} 

The classification of the spectra follows the Hanscom system, 
defined by \cite{kra02} for spectra from the Short-Wavelength 
Spectrometer (SWS) aboard the {\it Infrared Space Observatory 
(ISO)}, which is based partially on the classification of 
spectra of oxygen-rich AGB variables developed by 
\cite{sp95} for data from the Low-Resolution Spectrograph on 
the {\it Infrared Astronomical Satellite}.  \cite{slo08} 
explained how the method was modified for IRS data from {\it 
Spitzer}.  In the Hanscom system, spectra are divided into 
groups, based on their overall color, and one- or two-letter 
designations are added to describe the dominant spectral 
features.  All of our spectra fall into Group 1 (for blue 
spectra dominated by stellar continua and showing no obvious 
dust), Group 2 (for stars with dust), and Group 3 (for 
spectra dominated by warm dust emission).  Most of our 
spectra can be classified as either naked stars (``1.N'') or 
stars showing silicate emission (``2.SE.'').  

The classification depends primarily on two quantities.  
The first quantity is Dust Emission Contrast (DEC), defined 
as the ratio of the dust excess to the stellar continuum, 
integrated from 7.67 to 14.03~\mum.  For the stellar 
continuum, we assume a 3600 K Planck function, fit to the 
spectrum at 6.8--7.4~\mum.  \citet[][1998]{sp95} assumed an 
Engelke function with 15\% SiO absorption at 8~\mum.  
Switching to this continuum would systematically shift our
DEC measurements upward by $\sim$0.04.  Table~\ref{TblSp}
does not include this systematic error in the uncertainties,
although it should be kept in mind that the actual stellar
continuum could vary from one source to the next.  In this
sample and with these assumptions, a DEC of 0.10 separates 
those stars which we visually identify as naked from those 
with apparent dust excesses.

%NOTE - DECs in the table currently contain only the random
%error in the fit, not the systematic error from assuming
%different continua.  I will provide these, with a note, and
%update the figures below with larger error bars to reflect
%this uncertainty.

The second quantity, determined for the spectra with an
oxygen-rich dust excess, is the ratio of the excess emission 
at 11 and 12~\mum\ ($F_{11}/F_{12}$).  To measure this flux 
ratio, we follow the method of \cite{sp95}, measuring the 
excess at 10, 11, and 12~$\mu$m and plotting the flux ratio 
$F_{10}/F_{12}$ as a function of $F_{10}/F_{11}$.  
Figure~\ref{FigSDS} shows that all oxygen-rich sources fall 
on or close to the {\it silicate dust sequence}, which was 
defined as a power law:  $F_{10}/F_{12}$ = 1.32 
$(F_{10}/F_{11})^{1.77}$.  In the bottom left, the point 
furthest from the power law is the carbon star Lyng{\aa}~7~V1.
For the remaining sources, the flux ratio $F_{11}/F_{12}$ is 
determined by finding the point on the silicate dust sequence 
closest to the point defined by the measured ratios 
$F_{10}/F_{11}$ and $F_{10}/F_{12}$.  The corrected ratio 
defines the silicate emission (SE) index, which runs from 1 
for $F_{11}/F_{12} < 0.85$ to 8 for $F_{11}/F_{12} \ge 1.55$.  
Four sources have positions shifted to the right of the
silicate dust sequence in the region where 1.3 $<$
$F_{10}/F_{12}$ $<$ 1.8; from top to bottom, they are
NGC~6712~V7, NGC~6838~V1, NGC~6356~V5, and NGC~6760~V4.
Their spectra have typical silicate emission features, and
we suspect that the shift arises from a dust component in 
addition to the usual mixture of amorphous alumina and
amorphous silicates.

The corrected flux ratio $F_{11}/F_{12}$ quantifies the dust 
composition \citep{es01}.  Amorphous alumina dominates the
spectra with low flux ratios (or SE indices 1--3), while
amorphous silicates dominate the highest flux ratios 
(SE6--8).  For the remainder of the paper, we will use the
quanitity $F_{11}/F_{12}$ to distinguish the spectra by the
composition of their dust. 

% The intermediate region could arise from 
% combinations of the two, from limited self-absorption in the 
% 10~\mum\ silicate emission feature, or perhaps from 
% crystalline grains.  

Lyng{\aa}~7~V1 is classified in the Hanscom system as ``2.CE''
(carbon-rich emission).  Of the remainder, 10 are naked 
(``1.N''), and 28 have oxygen-rich dust.  Most are classified 
in the sequence from to ``2.SE1'' to ``2.SE8'', but seven
spectra require special attention, as explained next.

Strong silicate self-absorption can push spectra down 
the silicate dust sequence, as has happened for Palomar~6~V1.  
This source mimics a 2.SE1 spectrum in Figure~\ref{FigSDS}, 
but its spectrum (Figure~\ref{FigSp7}) clearly shows 
self-absorption at 10~\mum.  That, and the fact that the 
spectrum peaks past 15~\mum, result in a classification of 
``3.SB''.

Six of the spectra are distinctly different from the rest and
are examined more closely in \S~\ref{SecSX} and \ref{SecSY} 
below.  Three clearly show the presence of crystalline 
silicates in the 10-\mum\ feature; they are classified as 
``SX'' instead of ``SE''.  Three others show an unusual 
spectral emission feature peaking between 11 and 12~\mum.  
To distinguish them from the remainder, we will give them the 
new (and possibly temporary) classification of ``SY''.  Both 
the SX and SY sources are indexed identically to the SE 
sources, although the interpretation of that index may differ.

The classifications in Table~\ref{TblSp} include the suffixes 
`t', `f', and `x', indicating the presence of a 13-\mum\ 
feature, a 14-\mum\ feature, and longer-wavelength features 
from crystalline silicates, respectively.  These emission 
features and the criteria for their classification are 
treated in more detail below (\S~\ref{SecSX} and 
\ref{SecNarrow}).

\subsection{Mass-loss Rates} % Sec. 4.2
\label{SecMLR}

The primary objective of this project is to quantify how the 
rate of dust production by evolved oxygen-rich stars depends 
on metallicity.  The dust-production rate is identical to the 
dust mass-loss rate, which is the ratio of the overall 
mass-loss rate divided by the gas-to-dust ratio.  We use two 
methods to quantify the amount of dust in the circumstellar 
shell:  the DEC described above and the [7]$-$[15] color 
defined by \cite{slo08} and described below.  To relate these 
to the dust mass-loss rate, we use the set of radiative 
transfer models fitted to 86 evolved stars in the Magellanic 
Clouds by \cite{gro09}.  Their models determine a dust mass 
from the opacity needed to fit the IRS spectroscopy and 
overall spectral energy distribution, assume a constant 
outflow velocity (10 km s$^{-1}$) to determine the dust 
mass-loss rate, and assume a gas-to-dust ratio of 200 to 
determine the overall mass-loss rate.

The [7]$-$[15] color integrates the total flux density (from 
star and dust) in the wavelength intervals 6.8--7.4 and 
14.4--15.0~\mum.  \cite{gro09} recently showed that the 
[7]$-$[15] color tracks the mass-loss rate:
\begin{equation}
\log \dot{M} ({\rm M}_{\sun}/{\rm yr}) 
   = 1.759 ( [7]-[15] ) - 8.664.
\end{equation}
They also found that the DEC correlates with mass-loss rate
as well:
\begin{equation}
\log \dot{M} ({\rm M}_{\sun}/{\rm yr}) = 1.392 DEC - 6.484,
\end{equation}
for mass-loss rates less than 10$^{-5.5}$ M$_{\sun}$/yr.  At
higher rates, the shell becomes optically thick enough to
drive the 10~\mum\ silicate emission feature into 
self-absorption, and as a result, the DEC breaks down as a 
useful measure of dust content.

For all sources except the carbon star Lyng{\aa}~7~V1 
(\S~\ref{SecLynga7V1}) and three Cepheids not observed at 
15~\mum, we estimated mass-loss rates using Equations (1) and 
(2).  We began with [7]$-$[15] color.  For those sources 
where [7]$-$[15] $>$ 1.80, indicating a mass-loss rate above 
10$^{-5.5}$ M$_{\sun}$/yr, we used only Equation (1).  For 
those sources below this limit, we also estimated a mass-loss 
rate from the DEC and averaged the two.  We assumed a minimum 
DEC of 0.0 for the purposes of estimating mass-loss rates.

Table~\ref{TblSp} presents the results as overall mass-loss 
rates, in order to facilitate comparison to other published 
mass-loss rates.  To convert overall mass-loss rate to
dust-production rate, one needs to divide by the assumed 
gas-to-dust ratio of 200, or in log space, to subtract 2.30.  
For the remainder of the paper, we will focus on the 
dust-production rate, which avoids assuming a gas-to-dust
ratio.

\subsection{Dust Production Dependencies} % Sec. 4.3
\label{SecDPR}

\begin{figure*} % Fig. 13
%\includegraphics[width=6.5in]{figures/f3pc.eps}
%\begin{figure}
\includegraphics[width=6.5in]{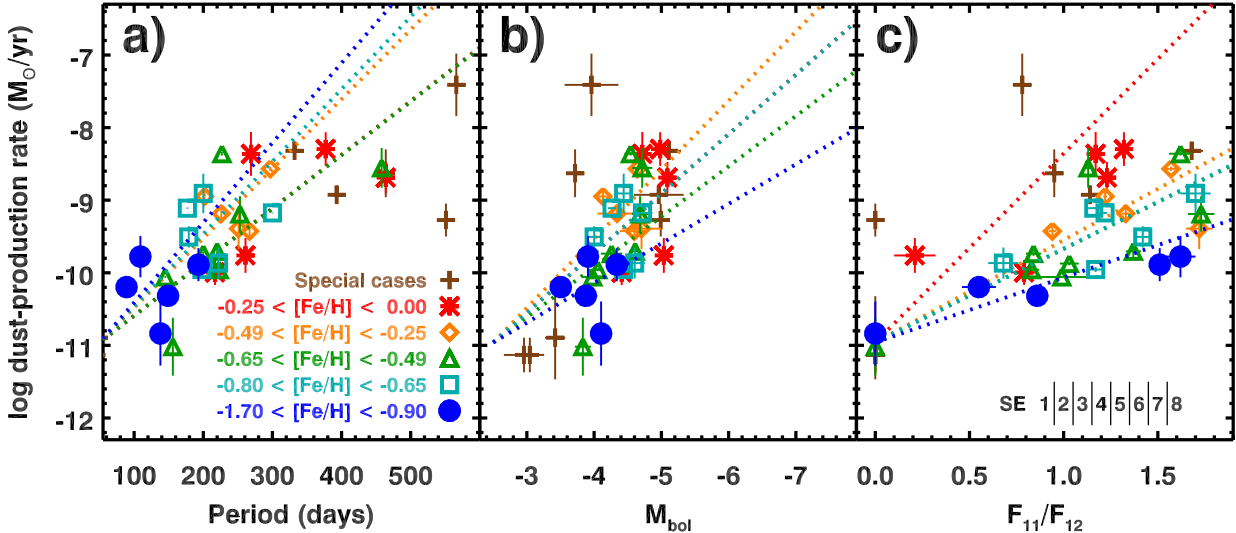}
\caption{The log of the dust-production rate (DPR) for each 
of the stars in our sample as a function of pulsation period 
(Panel a), bolometric magnitude (Panel b), and corrected flux 
ratio F$_{11}$/F$_{12}$ (Panel c).  The overall mass-loss 
rate is the product of the DPR and the gas-to-dust ratio,
which in Table~\ref{TblSp} is assumed to be 200 (a difference
of 2.30 in log space).  The corrected flux ratio in Panel c 
locates the position of a source along the silicate dust 
sequence (Fig.~\ref{FigSDS}) and measures the relative 
contributions from amorphous alumina (low F$_{11}$/F$_{12}$) 
and amorphous silicates (high F$_{11}$/F$_{12}$).  The dotted 
lines give the mean slope of each metallicity group, assuming 
a common y-intercept, as explained in the text.} 
\label{Fig3P}
\end{figure*} 
%\end{figure} 

\begin{figure} % Fig. 14
\includegraphics[width=3.5in]{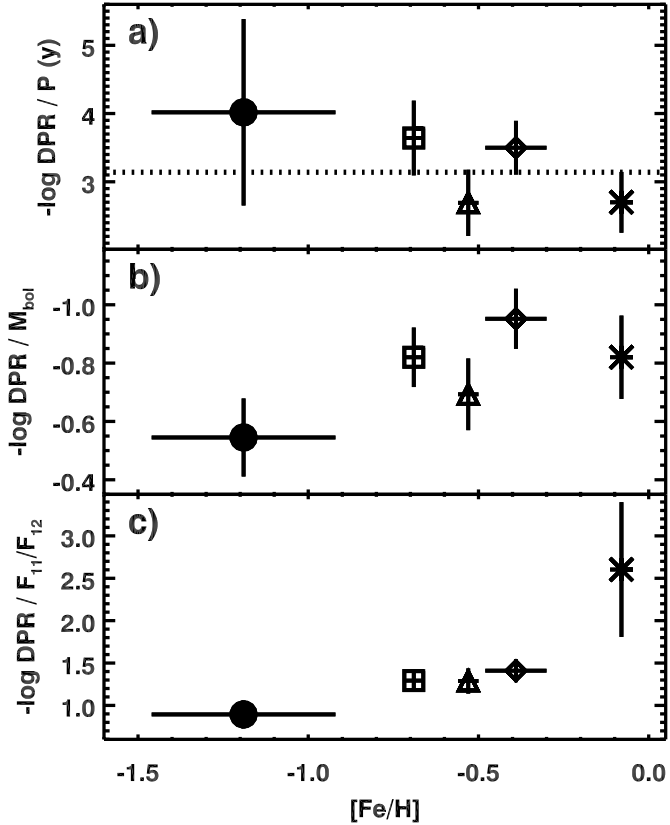}
\caption{The slopes from Fig.~\ref{Fig3P} plotted as a 
function of metallicity, with vertical error bars based on 
the uncertainty in the mean (standard deviation divided by 
the square root of the sample size.  The slopes are identical 
for overall mass-loss rate (MLR) and dust-production rate 
(DPR).  These plots show more clearly how the mass-loss rate 
depends on bolometric magnitude and position along the 
silicate dust sequence ($F_{11}/F_{12}$), but not on 
pulsation period (as indicated by the horizontal dotted 
line in Panel a).  Symbol shapes are as defined for 
Fig.~\ref{FigPM}, \ref{FigSDS}, and \ref{Fig3P}.}
\label{FigZ}
\end{figure}

Generally, the mass-loss rate and dust production rate will 
increase as a star evolves up the AGB, due to the reduced 
surface gravity and higher luminosity of the central star.  
This increase will dominate the subtler effect of 
metallicity.  To examine the role of metallicity, one would
ideally plot dust production vs.\ luminosity for samples
of different metallicities and compare them.  Past 
comparisons between evolved stars in the Galaxy and the 
Magellanic Clouds have been hampered by the poorly 
constrained distances to Galactic sources, and consequently, 
pulsation period has been used as a proxy for luminosity.  
Here, we can overcome this difficulty, since the distances to 
the globular clusters in our sample are known and we can 
determine bolometric magnitudes directly.

Figure~\ref{Fig3P} plots the derived mass-loss rates as a 
function of pulsation period (Panel a), bolometric magnitude 
(Panel b), and the corrected flux ratio $F_{11}/F_{12}$, 
which quantitifies the dust composition (Panel c).  In 
general, the rate of dust production clearly increases as the 
stars grow brighter, their pulsation periods increase, or the 
dust grows more silicate rich.  Panel c is analogous to 
Figure~4 by \cite{sp95}, which showed that alumina-rich 
dust (SE1--3) is seen only in low-contrast shells, while 
silicate-dominated dust (SE6--8) can exhibit a wide range of 
dust emission contrasts.  The globular sample follows the same 
trend.

To examine how the metallicity might influence the tendency
of the dust production to increase with increasing period, 
luminosity, or silicate/alumina dust ratio, we have 
determined a mean slope for each metallicity-defined 
subsample, including all the spectra depicted in 
Figures~\ref{FigSp1} through \ref{FigSp6}, assuming that they 
have same y-intercept (which we arbitrarily defined).  These 
slopes appear in Figure~\ref{Fig3P} as dotted lines.
Figure~\ref{FigZ} plots the slopes of these fitted lines as a
function of metallicity, showing how the metallicity modifies
the more obvious relations of dust production with period,
luminosity, and dust content.  The uncertainties in 
Figure~\ref{FigZ} are the formal uncertainties in the mean
\citep[the standard deviation divided by the square root of
the sample size;][]{bev69}.

Panel a in Figures~\ref{Fig3P} and \ref{FigZ} shows the
effect of metallicity when segregating the samples by period,
or more to the point, the lack of an effect.  While the data
may appear to show a negative relation, a horizontal line in 
Panel a of Figure~\ref{FigZ} can touch all of the error bars.  
This lack of a dependence on metallicity is a little 
suprising given that \cite{slo08} found a metallicity 
dependence in dust production when segregating by period and 
comparing oxygen-rich evolved stars in the Galaxy and the 
Magellanic Clouds.  However, the globular sample spans a 
wider range of metallicities, and the dependence of 
pulsation period on metallicity may be obscuring the 
dependence of dust production on metallicity.  One can see in 
Figure~\ref{Fig3P} that no star in the most metal-poor bin 
has a period greater than 200 days, while no star in the most 
metal-rich bin has a period less than 200 days.  The 
intermediate bins have intermediate periods.  \cite{woo90} 
noted that lowering the metallicity of an AGB star raises its 
effective temperature and thus reduces its radius, which also 
reduces its pulsation period.

Panel b in Figures~\ref{Fig3P} and \ref{FigZ} shows that 
segregating by bolometric magnitude reveals a slight 
dependence of dust production rates on metallicity, although
it is blurred somewhat by scatter in the sample.  

The impact of metallicity is clearest when segregating the
samples by dust content as measured by $F_{11}/F_{12}$ 
(Panel c in the two figures).  Whether the dust is 
alumina-rich or silicate-rich, the dust-production rate
increases as the metallicity increases.

Combining the evidence from the three methods of segregating
the globular sample, we conclude that metallicity is 
influencing the rate of dust production by AGB stars.  The
effect may be stronger than what we observe.  Implicit in
our calibration of dust-production rate from [7]$-$[15] and
DEC is the assumption by \cite{gro09} that the outflow
velocity is 10 km s$^{-1}$.  Recent CO obervations of carbon 
stars in the Galactic Halo by \cite{lag10} reveal a possible
dependence of the outflow velocity on metallicity.  While
this result requires confirmation, it is reasonable to 
consider the possibility that the outflow velocity increases 
with metallicity in our sample as well.  In this case, we 
would have to revise our dust-production rates upward for the 
higher metallicities, and the trends in Fig.~\ref{Fig3P} and 
\ref{FigZ} would be more apparent.

On the other hand, if the gas-to-dust ratio increases
at lower metallicity, this would decrease the dependency
for overall mass-loss rate.  The likelihood depends on how
important the role of dust is in the mass-loss process from
the AGB.

\subsection{Amorphous alumina} % Sec. 4.4
\label{SecAl}

Figure~\ref{FigSDS} and Figure~\ref{Fig3P} (Panel c) show 
that all of the metallicity bins contain spectra classified 
as SE1--3, which arise from shells dominated by amorphous 
alumina \citep{es01}.  We conclude that AGB stars can produce 
alumina-rich dust, regardless of their initial metallicity.

\cite{slo08} found little incidence of alumina-rich dust in 
their Magellanic samples, which they attributed to a possible 
underabundance of aluminum in more primitive stars, but our 
detection of alumina-rich dust shells at even lower 
metallicities repudiates that argument.  \cite{sp98} found 
that Galactic supergiants rarely produced alumina-rich dust 
shells.  Figure~\ref{FigHist} shows that the Magellanic 
sample studied by \cite{slo08} is more luminous, and it 
follows that it generally contains more massive stars.  The
lack of low-mass stars in the Magellanic samples may explain 
the missing alumina-rich dust.

%We need to plot F11/F12 vs. bolometric luminosity, or
%vice versa...

\subsection{Warm crystalline dust} % Sec. 4.5
\label{SecSX}

\begin{figure} % Fig. 15
\includegraphics[width=3.5in]{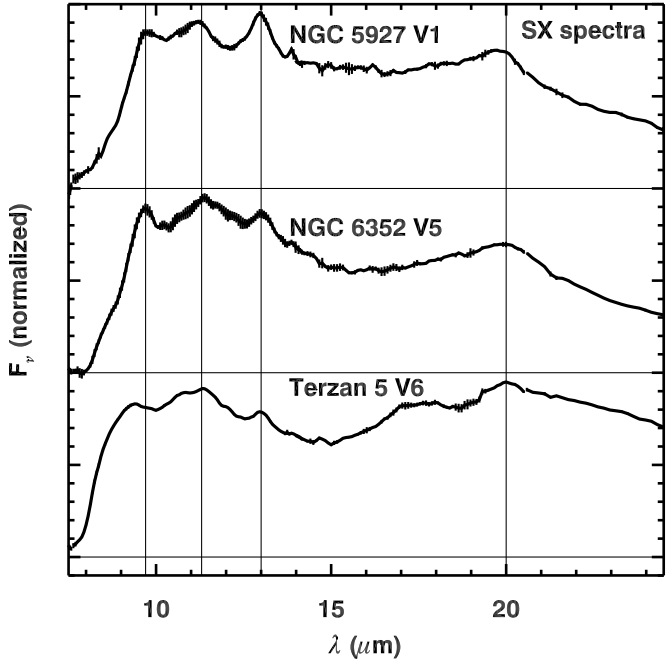}
\caption{The three spectra classified as ``SX'' due to the
likely presence of crystalline silicate emission at 10 and 
11~\mum.  These spectra are plotted after a stellar continuum
has been fitted from 6.8 to 7.4~\mum\ and removed.  All three 
show features at each of the wavelengths marked by vertical 
lines:  9.7, 11.3, 13, and 20~\mum.  Terzan~5~V6, with the 
``9.7''-\mum\ feature shifted to 9.3~\mum\ and a clear 
contribution from silicates at 18~\mum\ feature, may be 
affected by some self-absorption in the 10~\mum\ silicate 
feature.}
\label{FigSX}
\end{figure}

\begin{deluxetable}{lcc} % Table 5
\tablecolumns{3}
\tablewidth{0pt}
%\tabletypesize{\small}
\tablecaption{Narrow dust emission feature strengths\label{TblDust}}
\tablehead{
  \colhead{ } & \multicolumn{2}{c}{Feature strength/dust strength 
  (\%)\tablenotemark{a}} \\
  \colhead{Target} & \colhead{13 \mum} & \colhead{20 \mum} 
}
\startdata
 NGC 362 V2    & \nodata         & 0.70 $\pm$ 0.93 \\
 NGC 362 V16   & \nodata         & 0.60 $\pm$ 0.21 \\
 NGC 5139 V42  & \nodata         & 0.21 $\pm$ 0.08 \\
%NGC 5904 V84  &                 & \\
 NGC 5927 V1   & 1.19 $\pm$ 0.09 & 0.70 $\pm$ 0.05 \\
 NGC 5927 V3   & 0.11 $\pm$ 0.01 & 0.07 $\pm$ 0.03 \\
%Lynga 7 V1    &                 & \\
%NGC 6171 V1   &                 & \\
%NGC 6254 V2   &                 & \\
 NGC 6352 V5   & 0.46 $\pm$ 0.10 & 0.60 $\pm$ 0.04 \\
 NGC 6356 V1   & 0.01 $\pm$ 0.01 & \nodata         \\
 NGC 6356 V3   & \nodata         & 0.09 $\pm$ 0.24 \\
 NGC 6356 V4   & 0.61 $\pm$ 0.31 & 0.23 $\pm$ 0.27 \\
 NGC 6356 V5   & 0.28 $\pm$ 0.12 & \nodata         \\
%NGC 6388 V3   &                 & \\
%NGC 6388 V4   &                 & \\
 Palomar 6 V1  & \nodata         & 0.09 $\pm$ 0.03 \\
 Terzan 5 V2   & 0.24 $\pm$ 0.13 & \nodata         \\
 Terzan 5 V5   & 0.06 $\pm$ 0.07 & 0.13 $\pm$ 0.03 \\
 Terzan 5 V6   & 0.28 $\pm$ 0.02 & 0.49 $\pm$ 0.03 \\
 Terzan 5 V7   & 0.03 $\pm$ 0.01 & 0.16 $\pm$ 0.02 \\
 Terzan 5 V8   & 0.21 $\pm$ 0.11 & 0.80 $\pm$ 0.16 \\
 Terzan 5 V9   & 0.27 $\pm$ 0.03 & 0.35 $\pm$ 0.03 \\
 NGC 6441 V1   & 1.12 $\pm$ 0.20 & 0.93 $\pm$ 0.21 \\
%NGC 6441 V2   &                 & \\
 NGC 6553 V4   & 0.37 $\pm$ 0.06 & 0.24 $\pm$ 0.05 \\
 IC 1276 V1    & 0.45 $\pm$ 0.18 & 0.17 $\pm$ 0.06 \\
 IC 1276 V3    & \nodata         & 0.23 $\pm$ 0.05 \\
 Terzan 12 V1  & \nodata         & 0.08 $\pm$ 0.02 \\
%NGC 6626 V17  &                 & \\
 NGC 6637 V4   & \nodata         & 0.07 $\pm$ 0.02 \\
%NGC 6637 V5   &                 & \\
 NGC 6712 V2   & \nodata         & 0.04 $\pm$ 0.07 \\
%NGC 6712 V7   &                 & \\
 NGC 6760 V3   & \nodata         & 0.32 $\pm$ 0.08 \\
 NGC 6760 V4   & \nodata         & 0.10 $\pm$ 0.06 \\
%NGC 6779 V6   &                 & \\
 Palomar 10 V2 & 0.22 $\pm$ 0.01 & 0.05 $\pm$ 0.03 \\
 NGC 6838 V1   & 0.32 $\pm$ 0.15 & 0.32 $\pm$ 0.04 \\
\enddata
\tablenotetext{a}{Ratio of feature strength to total dust
emission from 5 to 35~\mum; see \S 4.4.}
\end{deluxetable}

\begin{deluxetable}{ccc} % Table 6
\tablecolumns{3}
\tablewidth{0pt}
\tablecaption{Wavelength intervals for extracting narrow dust 
              emission features\label{TblLam}}
\tablehead{
  \colhead{Feature (\mum)} &
  \multicolumn{2}{c}{Continuum intervals (\mum)} }
%  \colhead{$\lambda_{blue}$ (\mum)} & \colhead{$\lambda_{red}$ (\mum)} }
\startdata
13    & 12.20--12.35 & 13.50--13.65 \\
20    & 18.40--18.80 & 21.10--21.60 \\
\enddata
\end{deluxetable}

Figure~\ref{FigSX} shows the three spectra classified as 
``SX'', based on the splitting of the 10~\mum-emission 
feature from amorphous silicate grains into two components 
at 9.7 and 11.3~\mum.  \cite{slo06} found one source in the 
LMC which showed similar structure at 10~\mum, HV 2310, and 
they showed that an increased fraction of crystalline
silicate grains could explain the spectrum.  They suggested
that grains might form with more crystalline structure in
low-density dust-formation zones, as described below.
\cite{slo08} added a second 
source in the LMC:  HV 12667.

All three globular SX spectra also show a strong 13-\mum\ 
feature, as well as an additional component at 20~\mum.  
These features do not appear in the spectra of HV~2310 or 
HV~12667 in the LMC.  However, at least 24 other spectra in 
the globular sample show at least one of these features.  
Table~\ref{TblDust} presents the strengths of the features in 
the spectra where at least one was detected, expressed as a 
percentage of the total dust emission from 5 to 35~\mum.  We 
measured these strengths by fitting a line segment over the 
wavelength ranges given in Table~\ref{TblLam} and integrating 
in between.  We repeated the process using a spline to 
estimate the (dust) continuum, integrating over the same 
wavelength range.  Where both methods produced a continuum 
that followed the actual data to either side of the feature, 
we averaged the result.  The uncertainties in 
Table~\ref{TblDust} are the larger of the propagated error in 
the two extractions or the standard deviation between them.

\cite{slo03} found a correlation between the strength of the 
13- and 20-\mum\ features, quoting a Pearson correlation 
coefficient of 0.82.  We have calculated the same coefficient 
for all of the sources in the present sample to be 0.62, 
notably but understandably less given the lower S/N and 
spectral resolution of the current set of spectra.

Following \cite{slo03}, we assign a suffix of ``t'' to the 
classification of all sources with a 13~\mum\ feature 
stronger than 0.1\% of the total dust emission, provided that 
the S/N of the detection is 3.0 or more.  Where the feature 
exceeds 0.1\% of the total dust, but the S/N is 1.0--3.0, we 
classify the spectrum as ``t:''.  Eight spectra have clear 
13-\mum\ features, and six more have probable 13-\mum\ 
features, which accounts for 14 out of 30 spectra showing 
oxygen-rich dust emission, or 47\%.  Nearly all of the
variables in the sample are classified as Miras, and 
Figure~\ref{FigPK} reinforces this point.  \cite{slo96} 
estimated that only $\sim$20\% of Galactic Miras with 
oxygen-rich dust showed 13-\mum\ features; the percentage 
here is more than twice as high.  This difference might be 
due to the lower masses in the globular sample.

The specific grain which produces the 13-\mum\ feature 
remains an unsettled issue.  Whether it is crystalline 
alumina (corundum; Al$_2$O$_3$), as first proposed by 
\cite{gla94}, or spinel \citep[MgAl$_2$O$_4$;][]{pos99,fab01}, 
the groups favoring both candidates agree that an Al--O 
stretching mode produces the feature \citep[e.g.,][]{leb06}.  

Concentrating on this fact, we can infer that the presence 
of the 13-\mum\ feature indicates the presence of elemental 
aluminum.  The 14 spectra with probable 13-\mum\ features
are distributed evenly over all metallicities down to
[Fe/H]=$-$0.71.  Below this metallicity, all but two of
the stars are naked, limiting any possible conclusions for
that portion of the sample.  The presence of aluminum at all 
metallicities where it can be detected suports the conclusion
in \S~\ref{SecAl}:  The absence of alumina in the Magellanic 
samples was not a result of different abundances.  We see 
alumina-based dust species in the globular sample at any 
metallicity where we see oxygen-rich dust.

The difference between corundum vs.\ spinel as the carrier
of the 13-\mum\ features boils down to whether or not simple 
oxide grains like MgO (or perhaps FeO) are actively bonding 
with the alumina.  This difference is important in explaining 
the 20-\mum\ features, since \cite{slo03}, who favored 
corundum, suspected that the latter feature arises from 
crystalline silicates, while \cite{pos02} have proposed an 
origin in simple oxides ([Mg,Fe]O).  In the former case, the 
presence of the 13- and 20-\mum\ features may point to the 
presence of crystalline analogues to the amorphous alumina 
and silicates known to form in these dust shells, while in 
the latter, the two features may indicate a modified 
chemistry with enhanced oxides.  Thus, a resolution of the 
question would help us understand the pathways followed by 
the grain chemistry in these dust shells.

The presence of three spectra classified as ``SXt'' bolsters
the case for crystallinity as the origin of the 13- and 
20-\mum\ features, since we can now see an accompanying 
enhancement in crystalline olivine at 10~\mum.  These spectra 
are particularly intriguing, because they help to clarify the 
origin of the ``three-component spectra'', where narrow 
features at 11 and 13~\mum\ are superimposed on the broader 
10~\mum\ silicate emission feature \citep{lml88}.  Here, we 
can see the three components more clearly than before.

\cite{slo06}, following theoretical arguments by
\cite{gs98}, suggested that dust forming at lower densities 
may have a higher degree of crystallinity.  Each photon 
absorbed by a dust grain will nudge recently accumulated 
atoms across the surface.  If these atoms can be nudged 
enough times before the accumulation of an overlaying layer 
locks them into position, then they might fall into lowest
energy levels represented by the lattice structure.

A separate mechanism produces crystalline grains at higher
mass-loss rates.  The higher densities in the dust-formation
zone push the condensation temperature higher, and the grains 
can anneal into a crystalline structure before they cool 
\citep[see][and references therein]{fab00}.  Thus,
crystalline grains could form in cases of either particularly
high or particularly low mass-loss rates.

If indeed the appearance of features at 11, 13, 20, and 
28~\mum\ is due to enhanced crystallinity, one could argue
that this is evidence for a disk, which would retain the
grains close to the star long enough for them to be
radiatively annealed.  We believe this is unlikely for
most of the sources in our sample because disks would have 
higher optical depths and more cool dust than observed in 
these spectra.  Generally, self-annealing during formation
at low densities is a better explanation than disks.

Palomar~6~V1 may be a different matter.  If it is a cluster
member, its bolometric magnitude and period are inconsistent
with the evolutionary track defined by the AGB stars in this
sample (Figure~\ref{FigPM} and \S~\ref{SecSpecial}).  Its
spectrum is probably not produced by outflows from a normal
AGB star.  The presence of silicate absorption requires high
optical depth, which could indicate the presence of an
optically thick disk.  Grains in such a disk would remain
in the star's vicinity longer than in an outflow, giving
them time to be annealed, which could explain the presence of 
emission features from crystalline silicates at 23, 28, and 
33~\mum.

%However, we should emphasize that the various features
%involved, at 11, 13, 20, 28, and 32~\mum, do not rise and
%fall in unison.  While some of this variation could arise
%from noise in the data or differences in temperature, this
%behavior is consistent with the possibility that different
%mechanisms are contributing to the oberved spectra.  ...

\subsection{Unusual 11--12-\mum\ Emission Features} % Sec. 4.6.
\label{SecSY}

\begin{figure} % Fig. 16
\includegraphics[width=3.5in]{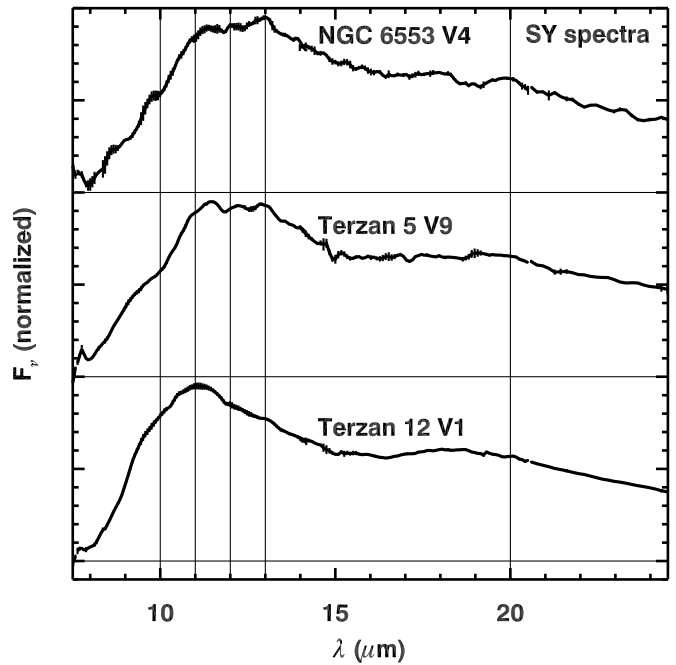}
\caption{The dust excess, after fitting and subtracting a
stellar continuum, from the three spectra classified as 
``SY'' due to their unusual 11--12~\mum\ emission features.
The vertical lines appear at wavelengths of 10, 11, 12, 13,
and 20~\mum.} \label{FigSY}
\end{figure}

Figure~\ref{FigSY} plots the three spectra classified as ``SY'', 
due to a previously unrecognized dust emission feature 
peaking in the vicinity of 11--12~\mum.  Two of the three
sources, NGC~6553~V4 and Terzan~5~V9, also show 13-\mum\ 
features along with a feature at 20~\mum.  Both spectra also 
show peaks in the emission at $\sim$11.5 and $\sim$12.2~\mum.  
Terzan~12~V1, while having general similarities to the other 
two spectra, differs in detail, with a single peak in the
dust emission at $\sim$11.1~\mum\ and a relatively 
normal accompanying 18~\mum\ feature.  

While the shape of the 11--12-\mum\ emission feature is 
unusual, the flux ratios at 10, 11, and 12~\mum\ place the 
three sources close to the silicate dust sequence in 
Figure~\ref{FigSDS} (in the SE1--2 range, with 
$F_{11}/F_{12} < 1.05$).  Normally, one would expect 
amorphous alumina to produce spectra with these flux ratios, 
but we have been unable to duplicate these spectra with 
optically thin dust shells consisting of only amorphous 
alumina, or even combinations of amorphous alumina and 
amorphous silicates.  Optically thick models remain untested.  
It may be worth recalling that one of the three sources, 
Terzan~5~V9, is something of an enigma (Figure~\ref{FigPM}), 
which could possibly be explained by the presence of a disk.

\subsection{Narrow Emission Features} % Sec. 4.7
\label{SecNarrow}

\begin{figure} % Fig. 17
\includegraphics[width=3.5in]{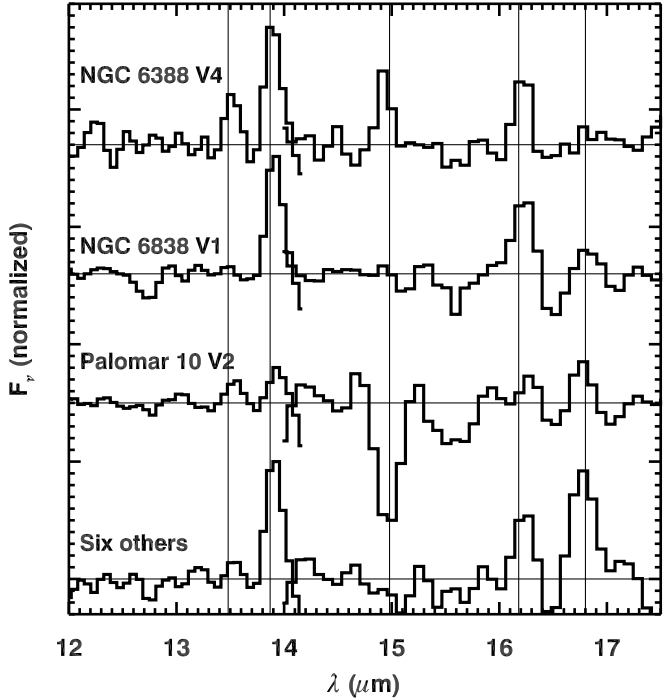}
\caption{Carbon dioxide emission bands in the spectra of 
three individual sources and the coaddition of six more.
A spline has been fitted and removed from each spectrum,
and the residual normalized to the maximum emission or
absorption.  The horizontal lines mark zero excess flux in 
each spectrum, and the vertical lines are at wavelengths 
of 13.48, 13.87, 14.98, 16.18, and 16.80~\mum.  The 
coadded spectrum at the bottom includes IC~1276~V1, 
NGC~5139~V42, NGC~5927~V1, NGC~6171~V1, NGC~6637~V4, and 
NGC~6760~V3.} \label{FigCO2}
\end{figure}

Many of the spectra show emission features at 14~\mum, but
the majority of these features are narrower than the 14-\mum\ 
features reported by \citet[][ 2008]{slo06}.  Only two 
sources show 14-\mum\ features with similar positions and 
widths to those in these previous papers:  NGC~6356~V1 and 
Palomar~6~V1.  They are classified as 2.SE8f and 3.SBfx, 
respectively.  The ``x'' for Palomar~6~V1 indicates the 
presence of emission from crystalline silicates at 23, 28, 
and 33~\mum.  The 14-\mum\ dust emission feature tends to be 
associated with strong amorphous silicate emission, as with 
NGC~6356~V1, or with crystalline features, as with 
Palomar~6~V1.

The narrow 14-\mum\ features seen in some spectra prevent a 
thorough search for the slightly broader 14-\mum\ feature 
detected in NGC~6356~V1 and Palomar~6~V1.  The narrow 
14-\mum\ features are centered close to 13.9~\mum, and some 
of the most pronounced examples are accompanied by a second 
emission feature at 16.2~\mum\ as well as absorption or 
emission at 15.0~\mum.  Figure~\ref{FigCO2} plots some 
examples.  

This combination of spectral features allows us to positively 
identify the bands as emission from CO$_2$.  \cite{jus98}
first identified these bands in spectra from AGB stars 
obtained with the SWS.  The detection of the same bands with 
the low-resolution modules of the IRS is a bit of a surprise, 
because its spectral resolution is almost an order of 
magnitude lower than the SWS.  $^{12}$CO$_2$ produces narrow 
emission bands at 13.87, 14.98, and 16.18~\mum, with the 
14.98~\mum\ band shifting into absorption in some cases 
\citep[see][]{cam00}.  Some spectra also show the emission 
band seen in SWS data at 13.48~\mum, as well as an 
additional band at 16.8~\mum, which \cite{slo03} argued was 
also from CO$_2$.  

% Speculative text from Mikako
% CO$_2$ main isotope normally has the strongest Q-branch 
% feature at 14.9\,$\mu$m, if the gas is in the LTE.
% However, the observed one usually do not show this feature 
% is not strongest, which is peculiar.
% This might suggests, that the feature might be self-absorbed, 
% or non-LTE effect.

The repeated appearance of features at the right wavelengths
in our sample is convincing, even if many of the individual
features in individual spectra remain in doubt.  Given the
noisy nature of the features, we have not attempted a 
quantitative analysis, though it is worth noting, that most 
of the sources included in Figure~\ref{FigCO2} are {\it not} 
13-\mum\ sources.  If this result held up with better spectra 
of globular cluster variables in the future, it would 
contradict the correlation found by \cite{slo03} for 
oxygen-rich AGB variables in the Galaxy, where CO$_2$ 
emission appeared in spectra also showing 13-\mum\ features.

% Fat 13 um features: NGC 5927 V1, NGC 6352 V5, NGC 6441 V1 (?),
% Terzan 5 V6 (?).

\subsection{Molecular absorption features} % Sec. 4.8
\label{SecAbs}

\begin{figure} % Fig. 18
\includegraphics[width=3.5in]{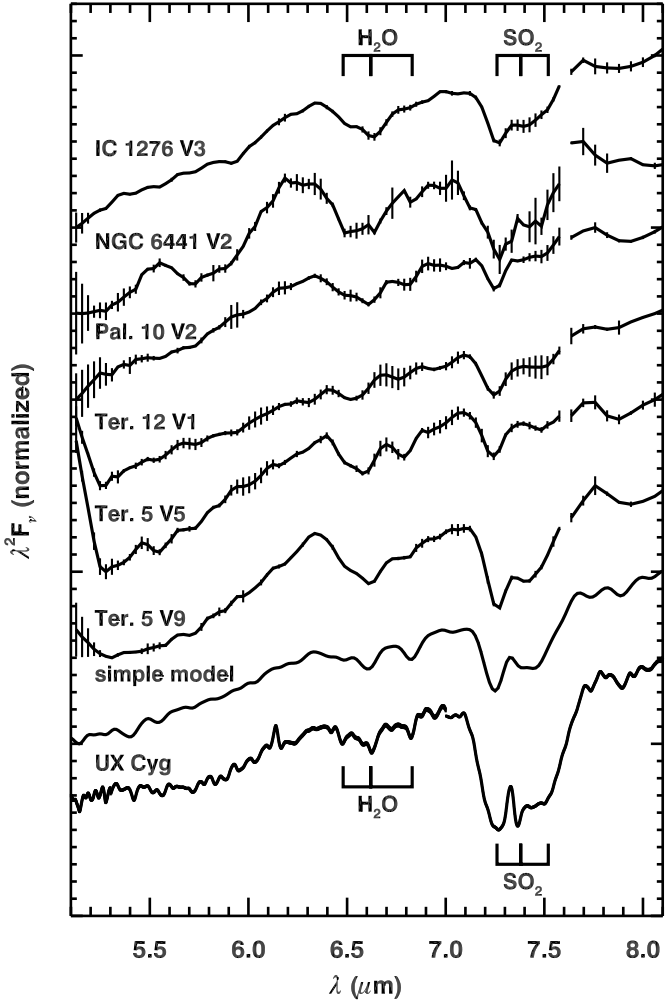}
\caption{Six IRS spectra showing SO$_2$ absorption bands
at 7.3--7.5~\mum, along with a simple model and a comparison 
spectrum of a Galactic AGB star, UX Cyg, obtained with the 
SWS on \iso.  The spectra are plotted in Rayleigh-Jeans units 
($\lambda^2$$F_{\nu}$ $\lambda^4$$F_{\lambda}$) so that the 
Rayleigh-Jeans tail of the blackbody function would be a 
horizontal line.  All of the spectra also show absorption 
from water vapor at 6.4--6.8~\mum.  The simple model
includes absorption from only H$_2$O and SO$_2$.} 
\label{FigSO2}
\end{figure}

\begin{figure} % Fig. 19
\includegraphics[width=3.5in]{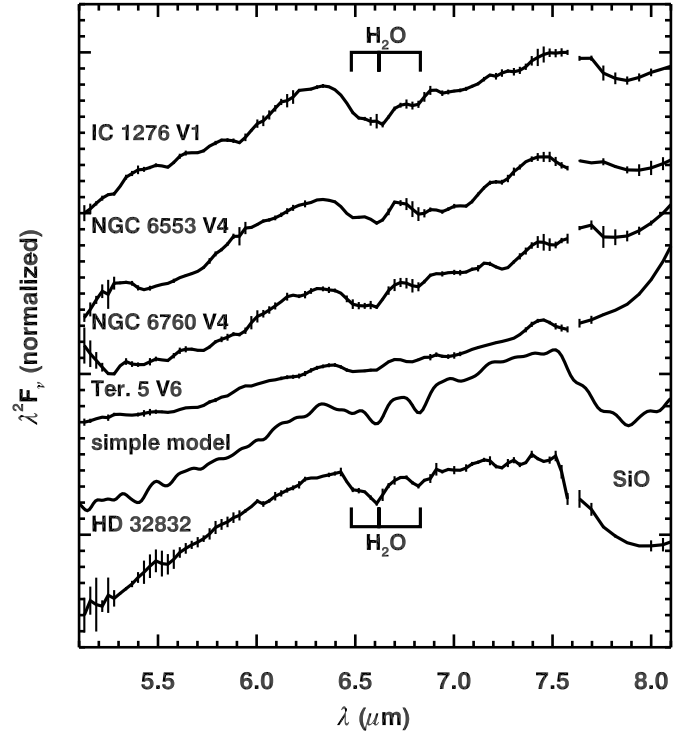}
\caption{Four IRS spectra of globular cluster variables 
showing water vapor absorption at 6.4--6.8~\mum, but no 
SO$_2$ absorption at 7.3--7.5~\mum, a simple model, and 
HD 32832, a naked Galactic comparison source with a spectral 
class of M4 III observed by the IRS.  The spectra are plotted 
in Rayleigh-Jeans units as in Fig.~\ref{FigSO2}.  The simple 
model only includes H$_2$O and SiO.} \label{FigAbs}
\end{figure}

\begin{figure} % Fig. 20
\includegraphics[width=3.5in]{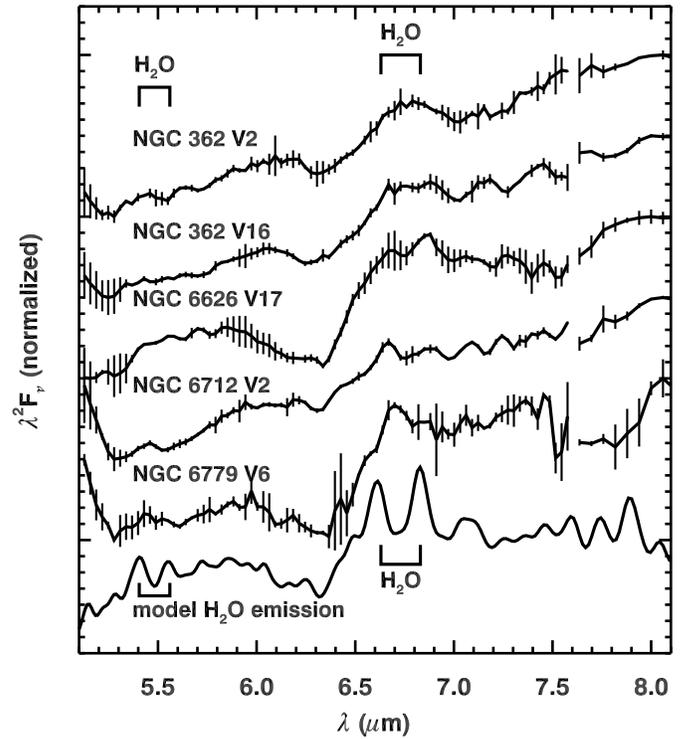}
\caption{Five IRS spectra showing a possible absorption
feature at $\sim$6.35~\mum.  The spectra are in 
Rayleigh-Jeans units, as in Fig.\ \ref{FigSO2} and 
\ref{FigAbs}.  The model shows water vapor in emission,
and it matches the structure seen in the data reasonably
well.} \label{FigEm}
\end{figure}

Many of the spectra in the globular sample show clear bands
from molecular gas shortward of 10~\mum.  The molecules
include SO$_2$ and H$_2$O.

Figure~\ref{FigSO2} presents six IRS spectra showing absorption 
from SO$_2$ at 7.3--7.5~\mum, along with an SWS spectrum of 
UX~Cyg, a Galactic AGB star and one of the first sources in 
which SO$_2$ was detected \citep{yam06}.  
SO$_2$ is the only sulphur-bearing molecule that has
been detected in oxygen-rich AGB stars.  One would not expect 
sulpher to be produced in low-mass AGB stars, and it would 
follow that the strength of the SO$_2$ band should depend on 
the initial metallicity of the star.  Thus, the presence of
this band in metal-poor clusters like IC~1276 and NGC~6441,
with [Fe/H] = $-$0.6 or less, is unexpected.

The seven IRS and SWS spectra in Figure~\ref{FigSO2} also 
show water vapor absorption at 6.4--6.8~\mum.  
Figure~\ref{FigAbs} presents four more globular spectra with 
water bands, but no SO$_2$, along with HD~32832, a Galactic M 
giant also observed with the IRS.  This comparison spectrum 
also shows strong SiO absorption, which is generally absent 
or weak in the globular sample.  

Figures \ref{FigSO2} and \ref{FigAbs} both contain synthetic 
absorption spectra based on plane-parallel radiative transfer 
models \citep{mat02a} and line lists from HITRAN \citep{rot09}.  
These models include only the main isotopes, and the updated 
line lists provide improved fits to the H$_2$O structure at
$\sim$6.3~\mum\ \citep{bar06}.  The excitation temperature 
of the molecules is 1800~K.  The absorption in the model in 
Figure~\ref{FigSO2} is due entirely to water vapor and 
SO$_2$.  In Figure~\ref{FigAbs}, it is from water vapor and 
SiO.
% and the column density is $1.2\times10^{21}$~cm$^{-2}$.

Figure~\ref{FigEm} presents five more IRS spectra showing an 
apparent absorption band with a peak opacity $\sim$6.35~\mum.
The figure also includes a model showing water vapor in 
emission, which duplicates reasonably well the structure in
the observed spectra.  Thus, the ``absorption'' is actually 
continuum between emission bands from H$_2$O to either side.  

\cite{mat02a} found that LPVs tend to show water vapor in 
emission near the maximum of their pulsation cycle, due to 
detached layers of water vapor in the extended atmosphere.  
Our five emission sources all have short periods, limiting
our certainty of the phase during the IRS observations.
The distinguishing characteristic of the emission sources is 
their periods.  All five sources in our sample with periods 
between 40 and 140 days show water-vapor emission.  The
variables with longer periods do not.

\cite{tsu01} detected water vapor in K giants, which are 
warmer than one would expect if the water vapor were in 
hydrostatic equilibrium.  Two of the emission sources in
our sample are Cepheids, which are even warmer, and we can
conclude that the water vapor around them is probably not in 
equilibrium.

% Water-vapor emission has been observed previously in only one 
% source, the red supergiant $\mu$~Cep \citep{tsu00}.  

There has been some unpublished speculation about a possible 
dust emission feature at 6~\mum\ and a possible correlation
with other features, such as the 14-\mum\ dust emission 
feature.  This ``6-\mum\ emission feature'' extends from 6.0 
to 6.5~\mum\ with a peak $\sim$6.2--6.3~\mum.  It is 
particularly noticeable in the spectra of NGC~6441~V2 and 
Terzan~5~V9 in Figure~\ref{FigSO2}.  However, the complex 
spectral structure in the 6-\mum\ region created by molecular 
absorption makes it more likely that this ``feature'' is really 
just the continuum between molecular absorption bands.

\section{Objects of note} % Sec. 5.0
\label{SecObject}

\subsection{Lyng{\aa}~7~V1} % Sec. 5.1
\label{SecLynga7V1}

The spectrum of Lyng{\aa}~7~V1 clearly identifies it as a
carbon star, and our analysis strongly supports its 
membership in the cluster (\ref{SecMemSum}).
While carbon stars in Galactic globular clusters are rare,
they are not unheard of.  \cite{wk98} listed three examples
in NGC~5139 ($\omega$ Cen) and noted that all three were CH
stars, which have probably become carbon-rich through mass
transfer in a binary system.  \cite{cot97} discovered a
candidate in the cluster M14; it too is a CH star.  
Lyng{\aa}~7~V1 may well be a similar object, which would 
explain how it could be a member of an old globular cluster 
and still be a carbon star.

For the sake of completeness, Table~\ref{TblSp} includes a 
mass-loss rate for Lyng{\aa}~7~V1 based on its [6.4]$-$[9.3] 
color, the relation of color to dust mass-loss rate defined 
by \cite{slo08}, and an assumed gas-to-dust ratio of 200.
Because of its potentially unusual evolutionary path, this
mass-loss rate might not be useful for intercomparisons
among samples with different metallicities.

\subsection{NGC~5139~V42} % Sec. 5.2
\label{Sec5139}

\cite{mcd09} obtained an N-band spectrum of NGC~5139~V42 
about four weeks before our IRS observation.  Like us, they 
observed little dust in the spectrum, while earlier 
mid-infrared photometry \citep[referenced by][]{mcd09} showed 
a clear excess at 10~\mum.  They concluded that this change 
in the measurements is likely real.  This spectral variation
emphasizes the temporal nature of these objects and the 
importance of obtaining statistically significant samples to 
average the variations out.  

\subsection{NGC~362~V2 and V16} % Sec. 5.3

Our classification of a star as ``naked'' means that it does
not show an identifiable excess in the 8--14~\mum\ region, 
based on its continuum level in the 5--8~\mum\ region.  
\cite{boy09b} recently detected a photometric excess in 
two of our ``naked'' sources, NGC~362~V2 and V16, based on 
photometry of the continuum at shorter wavelengths.  This 
excess is featureless (at the resolution of the photometry), 
leading them to identify amorphous carbon as a likely 
suspect.  Because the formation of CO would have left only 
oxygen to condense into dust in this oxygen-rich environment, 
the presence of carbon-rich dust would be most unexpected.
\cite{mcd10}, using the spectra presented here, show that
iron grains provide the best explanation for the featureless
excesses observed in these and other spectra.

\subsection{Cepheid variables} % Sec. 5.4
\label{SecCep}

Our globular sample includes four Cepheids.  All four are
type II Cepheids, with periods between $\sim$20 and 50~days, 
taken from the sample by \cite{mat06}.  They could be 
classified as W~Virginis or RV~Tauri stars, but the 
separation between these two groups is ambiguous especially 
among the globular cluster objects \citep[See][]{mat09}.

RV~Tau stars are often embedded within circumstellar dust
shells \citep{jur86}.  \cite{nc89} previously reported
an infrared excess from one source in our sample, 
NGC~6626~V17.  However, the current IRS spectrum does not
show a dust excess around this star, or the other three
Cepheids.  The difference may arise from temporal
variations.

It is interesting that the two Cepheids with longer periods,
NGC~6626~V17 and NGC~6779 V6, both show water-vapor emission
in their spectra.  The Galactic RV Tauri star R Scuti (period
147 days) also shows water-vapor emission \citep{mat02b}.

\section{Evolution on the AGB} % Sec. 6
\label{SecEvo}

\begin{figure} % Fig. 21
\includegraphics[width=3.5in]{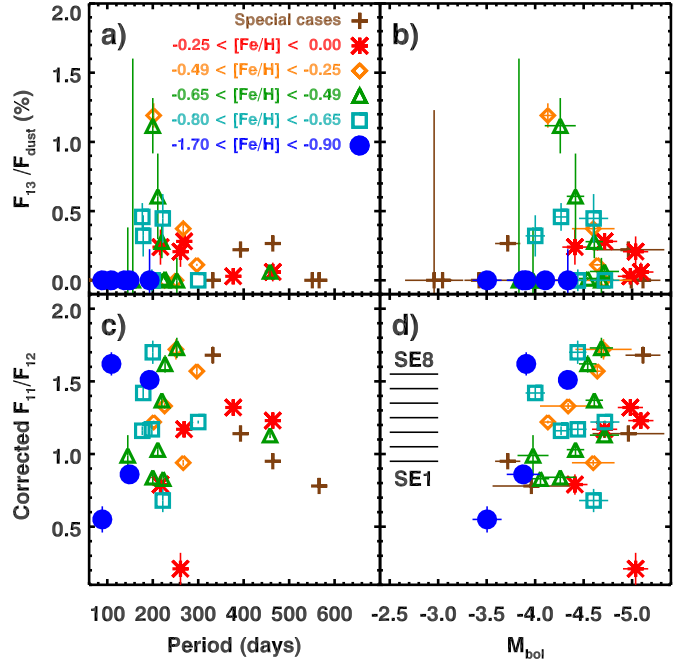}
\caption{The dependence of $F_{11}/F_{12}$ (dust composition) 
and the 13-\mum\ feature strength as a function of pulsation 
period and bolometric magnitude of the central star.  The 
flux ratio $F_{11}/F_{12}$ corresponds directly to the SE 
index, as indicated in Panel b.} 
\label{Fig4P}
\end{figure}

Figure~\ref{FigPM} compares the bolometric magnitudes and 
pulsation periods of the long-period variables in our 
globular cluster sample to theoretical evolutionary tracks
for low-mass AGB stars by \cite{vw93}.  The tracks start at 
the theoretical beginning of the thermally pulsing AGB, and 
they are roughly consistent with the data from our sample.  
All of the data appear to follow a mutually consistent 
evolutionary track, although that track does not follow the 
theoretical track for either 1.0-M$_{\sun}$ or 1.5-M$_{\sun}$ 
stars.  The two sources which fall to the right of and below 
the track are Terzan~5~V9 and Palomar~6~V1, and their 
location away from the rest of the data has already led us to 
conclude that either they are not actually cluster members, 
or they are not normal mass-losing AGB stars.

In their study of 47 Tuc, \cite{leb06} found that their
IRS spectra were consistent with the onset of dust formation 
at a luminosity of $\sim$2000~L$_{\sun}$, which corresponds 
to $M_{bol} \sim -3.5$, followed by a switch in pulsation
mode from an overtone to the fundamental.  The dust spectra 
in their sample showed strong 13-\mum\ features at this
stage, which then grew comparatively weaker as amorphous
silicates begin to dominate.  They emphasized the association
of a stronger 13-\mum\ feature with the fainter and less 
evolved stars in their sample.  Our study cannot address the 
question of a switch in pulsation mode because all of our 
long-period variables appear to be pulsating in the 
fundamental mode, which is consistent with all of our
data being brighter than $M_{bol} \sim -3.5$.

Figure~\ref{Fig4P} tracks the dust properties in our sample 
with pulsation period and bolometric magnitude.  The top
panels show that the 13-\mum\ feature is most pronounced for 
pulsators with periods around $\sim$200 days and bolometric 
magnitudes from $\sim$$-$3.5 to $-$4.0.  Thus, our globular 
sample generally conforms to the scenario described by 
\cite{leb06}, with the 13-\mum\ feature growing weaker and 
the contribution from amorphous silicates increasing as the 
star evolves along the AGB.  We do not see any evidence for 
sudden shifts in spectral properties, but then our sample
shows no shift from overtone to fundamental mode, as seen
in 47 Tuc.

The bottom panels of Figure~\ref{Fig4P} show little
dependence of the corrected flux ratio $F_{11}/F_{12}$ with
either period or bolometric luminosity.  We have noted
previously that Galactic supergiants and the luminous AGB
stars and supergiants observed in the Magellanic Clouds show
little sign of amorphous alumina in their spectra.  The
globular sample, though, is restricted to lower luminosities,
and within that range, amorphous alumina and amorphous 
silicates can dominate the shells with little apparent
dependence on either period or luminosity.

\section{Summary} % Sec. 7
\label{SecSum}

We have presented the infrared spectra of a sample of 39
variable stars in 23 globular clusters.  The sample includes
four Cepheid variables, none of which show dust in their
spectra, and 35 long-period variables, of which we can 
confirm 31 as members of their clusters.  These stars are
either naked, show emission from oxygen-rich dust,
or in one case, show emission from carbon-rich dust.  
The variety of dust species is impressive, and we see 
amorphous silicates, amorphous alumina, emission from
crystalline grains at several wavelengths, and three spectra
which we have not yet been able to characterize.  The spectra
are also rich in molecular features, including absorption 
from H$_2$O and SO$_2$, and emission from H$_2$O and CO$_2$.

The main objective of this paper is to probe how the
quantity and composition of the dust depend on metallicity.
Our sample shows dust emission over the range $-$0.97 $\le$
[Fe/H] $\le$ $-$0.08.  Across this range, stars with lower
metallicities generally have less dust in their circumstellar
shells.  This trend is most readily apparent when plotting
(dust) mass-loss rate as a function of the dust composition
(as quantified by the flux ratio $F_{11}/F_{12}$), but it
is also noticeable when plotting mass-loss rate vs.\
bolometric magnitude.  The lack of a metallicity trend vs.\
pulsation period may arise from dependencies of the period on 
metallicity.

We find contributions from alumina-rich dust at every 
metallicity at which we see silicates.  This result indicates
that the lack of alumina in the Magellanic samples published
thus far is most likely a selection effect, since those 
samples do not include objects as intrinsically faint as 
those published here, and the brighter objects tend to show
mostly silicates.

Three spectra have a 10~\mum-emission feature split into
components at 9.7 and 11.3~\mum, along with a strong 13-\mum\ 
feature.  The splitting at 10~\mum\ is reminicent of two
spectra in the LMC which \cite{slo08} fitted with 
crystalline analogues of the amorphous silicates which 
produce the broad 10-\mum\ feature.  These spectra support
the case that the 13-\mum\ feature arises from 
the crystalline analogue of the amorphous alumina seen in 
several other spectra.  
% Nonetheless, the carrier of the
% 13-\mum\ feature and the related feature at 20~\mum\ has
% still not been firmly identified.

This initial paper has only scratched the surface of a rich
set of spectroscopic data.  A careful mineralogical analysis 
of the dust features should help address some of the 
questions left unanswered here.  In addition, the spectra are 
rich in molecular features, both in the 5--8-\mum\ region and 
in the 13.5--17~\mum\ region.  Detailed modelling will give 
us insight about the behavior fo these molecules in 
metal-poor environments.

While {\it Spitzer} has exhausted its cryogens and the IRS
no longer operates, there is still a strong need for further
infrared spectra of evolved stars in globular clusters.  The
sample here is only large enough to point to trends of
fundamental importance.  The need for more data and larger
samples is clear.

\acknowledgements

We thank W.~E.\ Harris for helpful advice navigating the
available data on globular cluster properties, and I.\ 
McDonald at Manchester for his helpful comments.  The 
referee also contributed many useful suggestions and
questions.
These observations were made with the {\it Spitzer Space
Telescope}, which is operated by JPL, California Institute of
Technology under NASA contract 1407 and supported by NASA
through JPL (contract number 1257184).  This research has
made use of the SIMBAD and VIZIER databases, operated at the
Centre de Donn\'{e}es astronomiques de Strasbourg, and the
Infrared Science Archive at the Infrared Processing and
Analysis Center, which is operated by JPL.

\end{document}